\documentclass[aps,reprint,twocolumn,showpacs,floatfix,superscriptaddress]{revtex4-2}
\usepackage{amsmath,amssymb,color,braket,hyphenat,makeidx,xcolor}
\usepackage[ocgcolorlinks,colorlinks=true,linkcolor=blue,citecolor=red,linktocpage=true]{hyperref}

\usepackage{graphicx}
\usepackage[caption=false]{subfig}

\begin{document}
	%%%%%%%%%%%%%%%%%%%%%%%%%%%%%%%%%%%%%%%%%%%%%%%%%%%%%%%%%%%%%%%%%%%%%%%%%%%%%%%%%%%%%%%%%%%%%%%%%%%%
	\title{Quantum Entanglement Generation in Autonomous Thermal Machines: Effects of Non-Markovianity, Hilbert Space Structure, and Quantum Coherence}
	%%%%%%%%%%%%%%%%%%%%%%%%%%%%%%%%%%%%%%%%%%%%%%%%%%%%%%%%%%%%%%%%%%%%%%%%%%%%%%%%%%%%%%%%%%%%%%%%%%%%
	\author{A. Khoudiri}
\email{khoudiri.achraf@etu.uae.ac.ma}
\affiliation{Laboratory of R$\&$D in Engineering Sciences, Faculty of Sciences and Techniques Al-Hoceima, Abdelmalek Essaadi University, Tetouan, Morocco.}
\affiliation{
	The UM6P Vanguard Center, Mohammed VI Polytechnic University (UM6P),\\
	Rocade Rabat-Salé, Technopolis, 11103, Morocco.
}
%%%%%%%%%%%%%%%%%%%%%%%%%%%%%%%%%%%%%%%%%%%%%%%%%%%%%%%%%%%%%%%%%%%%%%%%%%%%%%%%%%%%%%%%%%%%%%%%%%%%

\author{K. El Anouz}
\email{kelanouz@uae.ac.ma}
\affiliation{Laboratory of R$\&$D in Engineering Sciences, Faculty of Sciences and Techniques  Al-Hoceima, Abdelmalek Essaadi University, Tetouan,	Morocco.}
%%%%%%%%%%%%%%%%%%%%%%%%%%%%%%%%%%%%%%%%%%%%%%%%%%%%%%%%%%%%%%%%%%%%%%%%%%%%%%%%%%%%%%%%%%%%%%%%%%%%

\author{A. El Allati}
\email{eabderrahim@uae.ac.ma}
\affiliation{Laboratory of R$\&$D in Engineering Sciences, Faculty of Sciences and Techniques Al-Hoceima, Abdelmalek Essaadi University, Tetouan,	Morocco.}

\affiliation{Université Grenoble Alpes, CNRS, LPMMC, 38000 Grenoble, France.}

	%%%%%%%%%%%%%%%%%%%%%%%%%%%%%%%%%%%%%%%%%%%%%%%%%%%%%%%%%%%%%%%%%%%%%%%%%%%%%%%%%%%%%%%%%%%%%%%%%%%%
	\begin{abstract}
We present a theoretical investigation of entanglement generation in an external quantum system during its interaction with a quantum autonomous thermal machine (QATM) under non-Markovian dynamics.  
		The QATM, composed of two qubits each coupled to independent thermal reservoirs, interacts with an external system of two additional qubits.  By analyzing the Hilbert space structure, energy level configurations, and temperature gradients, we define a common interaction between the QATM qubits and the external system qubits, which allows the definition of two thermodynamic cycles (A and B) governed by virtual temperatures and energy-conserving transitions.  
		We demonstrate that the QATM can act as a structured reservoir able to induce non-Markovian memory effects highlighted by negative entropy production rates.  Using mutual information and concurrence, we show that entanglement is generated only using cycle A, which is associated with stronger non-Markovian behavior and higher coherence correlations.  Our results demonstrate that temperature differences, Hilbert space structure, and coherence serve as quantum resources for controlling and enhancing entanglement in quantum thermodynamic settings, with parameters consistent with experimental superconducting qubit platforms.

	\end{abstract}
	\maketitle
	%---------------------------------------------------------------------------------
	\section{Introduction}
In the framework of quantum thermodynamics, quantum autonomous thermal machines (QATMs) provide challenging systems used for connecting quantum thermodynamic theory with emerging quantum technologies~\cite{INTRO1,INTRO2,INTRO3}. In addition, the interplay between quantum thermodynamics and quantum information plays a crucial role, such that quantum correlations can act as valuable resources in quantum thermodynamic processes~\cite{INTRO1A,INTRO2A}. In this context, several studies report apparent violations of the second law of thermodynamics and the Landauer bound due to correlation exchange between a system and its environment, which can be interpreted as an energetic contribution~\cite{INTRO3A}. Importantly, quantum correlations have been emphasized as thermodynamic quantum resources for information-processing tasks~\cite{INTRO4A}.

The role of QATMs has been explored in various contexts, particularly in enhancing the performance of quantum batteries and enabling low-temperature measurements through quantum thermometry~\cite{INTRO4,INTRO5}. Moreover, the relationship between the generation of quantum correlations using many-body systems and QATMs is investigated in Ref.~\cite{INTRO6}. However, several studies examined entanglement generation using QATMs, focusing on the correlation-driven thermal machine and its relationship with the second law of quantum thermodynamics, including the production of maximal steady-state entanglement~\cite{INTRO7,INTRO8,INTRO8A,INTRO8B}. Furthermore, the use of linear quantum thermal machines at low temperatures can generate entanglement between environments~\cite{INTRO9}. Besides, it has been shown that linear quantum thermal machines extend the laws of quantum thermodynamics to general autonomous quantum systems without external time-dependent Hamiltonians~\cite{INTRO9E,INTRO9F}.

The Hilbert space structure as a quantum resource in multilevel quantum thermal machines has been investigated in Ref.~\cite{INTRO9A}, where the relation between the Hilbert space dimension and the performance of the machine was analyzed. Related concepts such as virtual qubits and virtual temperatures were introduced to describe temperature reservoirs as quantum resources that ensure the autonomous operation of thermal machines and clarify their thermodynamic functioning~\cite{INTRO9B,INTRO9C}.

In our recent work~\cite{INTRO10}, we analyzed the dynamical influence of QATMs on an external system and showed that correlation exchange between the machine and the external system can generate memory effects. We also investigated the relation between non-Markovianity and the Landauer bound in the framework of autonomous quantum thermal machines~\cite{INTRO11}. These results further illustrate the connection between quantum thermodynamics and quantum information in autonomous thermal machines.

Motivated by these developments, we investigate here the generation of entanglement mediated by a QATM, extending our previous results~\cite{INTRO10}. In particular, we analyze the interplay between Hilbert space structure~\cite{INTRO11,INTRO12,INTRO13}, non-Markovianity~\cite{INTRO14,INTRO15,INTRO16,INTRO17}, virtual temperatures, and correlations in the entanglement generation process~\cite{INTRO18,INTRO19,INTRO20,INTRO21}.

In our model, the QATM consists of two qubits~\cite{INTRO9B,INTRO10}, each interacting with a bosonic reservoir at different temperatures (hot and cold). A structured external system composed of two qubits is coupled to the machine. The Hilbert space structure of the QATM and the external system is used to design the interaction under temperature and energy constraints~\cite{INTRO9A}. Using the concepts of virtual qubits and virtual temperatures~\cite{INTRO9C}, we define two thermodynamic cycles, denoted $A$ and $B$, allowing us to investigate the effects of reservoir temperature differences and Hilbert space structure.

The transitions between cycles $A$ and $B$ are expressed as functions of the energy spacing of the QATM qubits. The validity of the theoretical framework is examined numerically through the heat and temperature dynamics of each subsystem~\cite{INTRO22}. To ensure thermodynamic consistency, we analyze the second law of quantum thermodynamics via entropy production~\cite{INTRO23}, which characterizes irreversibility, and via the entropy production rate~\cite{INTRO24}, which quantifies information loss from the QATM and external system to the reservoirs. We observe a nonmonotonic entropy production rate that can become negative during the evolution, indicating non-Markovian dynamics, consistent with the findings of Popovic \textit{et al.}~\cite{INTRO25}.

This motivates the investigation of memory effects associated with cycles $A$ and $B$. Non-Markovianity is quantified using the measure introduced by Breuer \textit{et al.}~\cite{INTRO26}. Even with Markovian reservoirs, non-Markovian dynamics arises due to correlation exchange between the QATM and the external system~\cite{INTRO10,INTRO25}. To analyze entanglement generation, we evaluate total correlations between the external system qubits using mutual information~\cite{INTRO27} and quantify entanglement through concurrence~\cite{INTRO28}. While total correlations are similar in both cycles, entanglement is generated only when the QATM operates in cycle $A$, reflecting the different irreversibility and memory effects of the two cycles. We also investigate the role of correlation coherence~\cite{INTRO29} and local coherences~\cite{INTRO30}, showing that coherence acts as a resource for generating quantum correlations over time.

From an experimental perspective, the present work focuses on the theoretical framework. Nevertheless, all parameters are chosen within realistic regimes compatible with current superconducting-qubit implementations~\cite{INTRO31,INTRO32,INTRO33}.

The paper is organized as follows. In Sec.~\ref{sec:Theoretical Framework, Operation Cycle of the Thermal Machine, and Temperature and Energy Constraints}, we present the theoretical model, including the Hamiltonian description, operation cycles, and temperature and energy constraints. The Hamiltonian framework and system dynamics are discussed in Subsec.~\ref{sec:Hamiltonian Framework and System Dynamics}, while the machine cycles are introduced in Subsec.~\ref{sec:Quantum Thermal Machine Cycles}. In Sec.~\ref{sec:Dynamics of Quantum Thermodynamic Quantities: Heat and Temperature}, we analyze the thermodynamic dynamics. Heat dynamics are discussed in Subsec.~\ref{sec:Heat dynamics}, while temperature dynamics, entropy production, and entropy production rate are analyzed in Subsec.~\ref{sec:Temperature dynamics} and Subsec.~\ref{sec:Production entropy and production entropy rate}. Memory effects and their impact on entanglement generation are investigated in Sec.~\ref{sec:Impact of Memory Effects on Entanglement Generation in QATM Cycles for External System Qubits}. Non-Markovianity is discussed in Subsec.~\ref{sec:Non-Markovianity and correlation exchanges between the QATM and the external system}, while entanglement generation and the role of correlation coherence are analyzed in Subsec.~\ref{sec:Entangelment generaation} and Subsec.~\ref{sec:Role of coherence correlation on entanglement generation}. Finally, we conclude the paper.
	%---------------------------------------------------------------------------------

\section{Theoretical Framework, Operation Cycle of the Thermal Machine, and Temperature and Energy Constraints}
\label{sec:Theoretical Framework, Operation Cycle of the Thermal Machine, and Temperature and Energy Constraints}
\begin{figure}[ht!]
	\centering
	\includegraphics[scale=0.55]{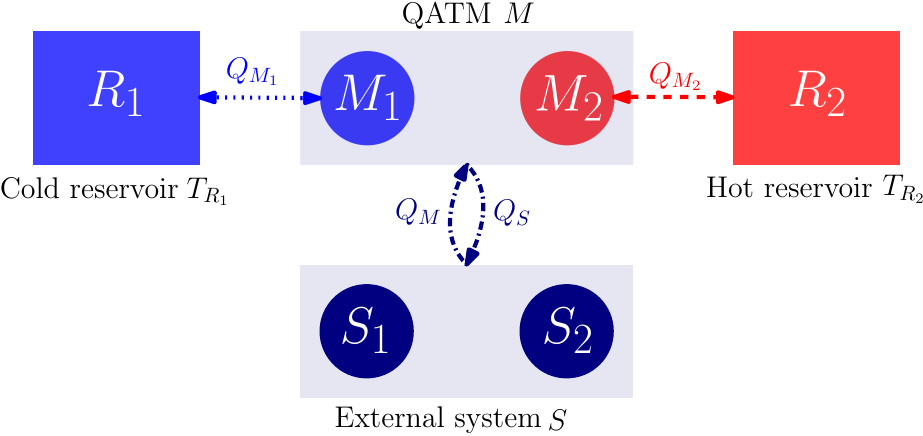}
    \caption{Schematic diagram of the quantum autonomous thermal machine (QATM) $M$ coupled to an external system $S$. The qubits $M_i$ and $S_i$ ($i=1,2$) denote the QATM qubits and the external system qubits, respectively. The heat fluxes $Q_M$ and $Q_S$ describe the energy exchange between $M$ and $S$, satisfying $Q_M = -Q_S$. The bosonic reservoirs $R_1$ (cold) and $R_2$ (hot) are coupled to the QATM qubits $M_1$ and $M_2$, respectively. Their temperatures are denoted by $T_{R_1}$ and $T_{R_2}$, with $T_{R_1} \leq T_{R_2}$. The quantities $Q_{M_i}$ ($i=1,2$) represent the heat fluxes between each reservoir $R_i$ and the corresponding qubit $M_i$.}
	\label{MODELREV}
\end{figure}

Our theoretical model consists of a quantum autonomous thermal machine (QATM), denoted by $M$, represented by two qubits $M_1$ and $M_2$. Each qubit is coupled, at thermal equilibrium, to a Markovian bosonic reservoir, respectively denoted by $R_1$ and $R_2$, at temperatures $T_{R_1}$ and $T_{R_2}$. The QATM $M$ is coupled to an external system $S$, represented by two qubits denoted by $S_1$ and $S_2$, respectively, as shown in Fig.~\ref{MODELREV}. 

We investigate the Hilbert space structure of the QATM $M$ and the external system $S$, as well as the temperature difference between $T_{R_1}$ and $T_{R_2}$, to realize thermodynamic cycles in the QATM. In this way, we explore how to enhance the memory effects in the external system $S$, and improve the generation and control of entanglement.

\subsection{Hamiltonian Framework and System Dynamics}
\label{sec:Hamiltonian Framework and System Dynamics}
The total Hamiltonian of our theoretical model is written in the following compact form:
\begin{equation}\label{TOTALH}
	\hat{H}=\hat{H}_{M} + \hat{H}_{S} + \hat{H}_{R} + \hat{H}_{MR} + \hat{H}_{MS},
\end{equation}
where $\hat{H}_{m}$, for $m=\{M,S,R\}$, denotes the free Hamiltonian of the QATM $M$, the external system $S$, and the reservoirs $R$ which are $R_1$ and $R_2$, respectively. Specifically,
\begin{align}
	\hat{H}_{M}&=\sum_{i=1}^{2}\hat{H}_{M_i},\quad \hat{H}_{M_i}=E_{M_i}\hat{\sigma}_{M_i}^{+}\hat{\sigma}_{M_i}^-,\nonumber\\
	\hat{H}_{S}&=\sum_{i=1}^{2}\hat{H}_{S_i},\quad \hat{H}_{S_i}=E_{S_i}\hat{\sigma}_{S_i}^{+}\hat{\sigma}_{S_i}^-,\nonumber\\	
	\hat{H}_{R}&=\sum_{i=1}^{2}\hat{H}_{R_i},\quad \hat{H}_{R_i}=\sum_{k}E_{R_{ik}}\hat{a}_{R_{ik}}^{\dagger}\hat{a}_{R_{ik}}.\nonumber
\end{align}
Here, $E_{M_i}$ and $E_{S_i}$ for $i=\{1,2\}$ are the energy level spacings of the QATM qubits $M_1(M_2)$ and the external system qubits $S_1(S_2)$, respectively. $E_{R_{ik}}$ denotes the energy of the reservoir $R_i$ corresponding to each mode $k$ of the set of bosonic harmonic oscillators.

The raising and lowering operators of the QATM and the external system qubits $M_i$ and $S_i$ are defined as $\hat{\sigma}_{M_i(S_i)}^- = \ket{0}_{M_i(S_i)}\bra{1}$ and $\hat{\sigma}_{M_i(S_i)}^{+} = \ket{1}_{M_i(S_i)}\bra{0}$. The operators $\hat{a}_{R_{ik}}$ ($\hat{a}_{R_{ik}}^{\dagger}$) are the annihilation and creation operators of the reservoir $R_i$ for the harmonic oscillator mode $k$.

The interaction between each reservoir $R_1$ ($R_2$) and the QATM qubits $M_1$ ($M_2$) is described, under the rotating-wave approximation ($E_{M_i} = E_{R_{ik}}$), by the interaction Hamiltonian $\hat{H}_{MR}$, defined as
\begin{align}\label{RM}
	\hat{H}_{MR}=\sum_{i=1}^{2}\sum_{k}g_{ik}(\hat{\sigma}_{M_i}\hat{a}_{ik}^{\dagger}+ H.c.).
\end{align}
Here, $g_{ik}$ is the coupling strength between the reservoir $R_i$ and the qubit $M_i$ for $i = \{1,2\}$.

The Hamiltonian $\hat{H}_{MS}$ represents the interaction between the QATM $M$ qubits $M_1$ and $M_2$ and the external system $S$ qubits $S_1$ and $S_2$, and is described by
\begin{equation}
	\hat{H}_{MS} = g \left( \ket{0_{M_1}1_{M_2}1_{S_1}0_{S_2}}\bra{1_{M_1}0_{M_2}0_{S_1}1_{S_2}} + H.c. \right),
\end{equation}
where $g$ denotes the coupling coefficient between the QATM $M$ and the external system $S$.

The interaction Hamiltonian $\hat{H}_{MS}$ induces the transitions
$\ket{0_{M_1}1_{M_2}} \otimes \ket{1_{S_1}0_{S_2}} \leftrightarrow \ket{1_{M_1}0_{M_2}} \otimes \ket{0_{S_1}1_{S_2}}$
under the condition of energy conservation~\cite{INTRO9A},
\begin{equation}
	E_{M_2} - E_{M_1} = E_{S_2} - E_{S_1}, \label{ConditionAB}
\end{equation}
The resonance condition between $M$ and $S$ drives the transitions in $M$,
$\ket{0_{M_1}1_{M_2}} \leftrightarrow \ket{1_{M_1}0_{M_2}}$, and in $S$,
$\ket{1_{S_1}0_{S_2}} \leftrightarrow \ket{0_{S_1}1_{S_2}}$,
through the exchange of an energy quantum $E_{M} = E_{S} = E_{M_2} - E_{M_1}$,
with $E_{M_2} > E_{M_1}$ and $E_{S_2} > E_{S_1}$, ensuring energy conservation during these transitions and satisfying the first principle of quantum thermodynamics~\cite{INTRO1,INTRO2}.
\subsection{Dynamics}	
The initial state of our total theoretical model, denoted by $\hat{\rho}(0)$, is defined as follows:
\begin{align}
	\hat{\rho}(0)=\hat{\rho}_{R_1-M_1}(0)\otimes \hat{\rho}_{R_2-M_2}(0) \otimes \hat{\rho}_{S}(0).
\end{align}
Initially, we assume no correlations between the subsystems, and $\hat{\rho}_{R_i-M_i}(0) = \hat{\rho}_{R_i}(0) \otimes \hat{\rho}_{M_i}(0)$ for $i = \{1,2\}$, where $\hat{\rho}_{R_i}(0)$ and $\hat{\rho}_{M_i}(0)$ are the initial states of the subsystems $R_i$ and $M_i$, respectively. The initial state of the external system is given by $\hat{\rho}_{S}(0) = \hat{\rho}_{S_1}(0) \otimes \hat{\rho}_{S_2}(0)$.

Under the weak interaction between the QATM $M$ qubits and the reservoirs $R_1$ and $R_2$, the total system evolves as
$\hat{\rho}(t) = \hat{U}(t)\hat{\rho}(0)\hat{U}^{\dagger}(t)$,
where $\hat{U}(t) = \exp(-i\hat{H}t)$ (with $\hbar = 1$).
The reduced state of the subsystems $M$ and $S$ is then obtained as
\begin{align}
	\hat{\rho}_{MS}(t) = \mathrm{Tr}_{R_1,R_2}\{\hat{\rho}(t)\}.
\end{align}
Hence, one can use the local standard Born--Markov master equation in Lindblad form~\cite{R6,R7}:
\begin{equation}\label{MASTER_EQUATION}
	\frac{d}{d t} \hat{\rho}_{MS}(t)=-i\left[\hat{H}_{S}+\hat{H}_{M},\hat{\rho}_{MS}(t)\right]+ \mathcal{L}[\hat{\rho}_{MS}(t)].
\end{equation}
Here, $-i\left[\hat{H}_{S} + \hat{H}_{M}, \hat{\rho}_{MS}(t)\right]$, describing the free evolution of the subsystems $M$ and the external system $S$, represents the reversible part of the dynamics of our theoretical model, while $\mathcal{L}[\hat{\rho}_{MS}(t)]$ is the dissipative (irreversible) part describing the decoherence effects of the reservoirs $R_1$ and $R_2$ on $M$ and $S$, defined as
\begin{align}
	\mathcal{L}[\hat{\rho}_{MS}(t)]=&-i\left[\hat{H}_{MS}, \hat{\rho}_{MS}(t)\right]
	+\sum_{i=1}^{2}\mathcal{D}^{[T_{M_i}]}[\hat{\rho}_{MS}(t)],\\
	\mathcal{D}^{[T_{M_i}]}[\hat{\rho}_{MS}(t)]=&\gamma_{i}(\bar{n}_{i}(T_{M_i},E_{M_i})+1)\mathcal{D}^{[\hat{\sigma}_{M_{i}}^-]}[\hat{\rho}_{MS}(t)] \nonumber\\
	&+\gamma_{i}\bar{n}_{i}(T_{M_i},E_{M_i})\mathcal{D}^{[\hat{\sigma}_{M_{i}}^{+}]}[\hat{\rho}_{MS}(t)],\\
	\mathcal{D}^{[\hat{\sigma}_{M_{i}}^{\pm}]}[\hat{\rho}_{MS}(t)]&=\hat{\sigma}_{M_{i}}^{\pm}\hat{\rho}_{MS}(t)\hat{\sigma}_{M_{i}}^\mp-\frac{1}{2}\{\hat{\sigma}_{M_{i}}^\mp\hat{\sigma}_{M_{i}}^{\pm},\hat{\rho}_{MS}(t)\},\nonumber\\
\end{align}

Here, $\gamma_{i}$ denotes the corresponding dissipation rate, and $\bar{n}_{i}(T_{M_i},E_{M_i})$ represents the average number of harmonic oscillators in the thermal bosonic reservoir $R_i$ at temperature $T_{M_i}$ corresponding to the energy gap $E_{M_i}$, defined as (with $\hbar = 1$)
\begin{equation}
	\bar{n}(T_{M_i},E_{M_i})=\frac{1}{\exp\left(\frac{E_{M_i}}{T_{M_i}}\right)-1}, \quad \text{for } i=\{1,2\}.
\end{equation}

It is noted that the coupling between $M_i$ and its corresponding reservoir $R_i$ ($i = \{1,2\}$) in the Hamiltonian of Eq.~\ref{RM} satisfies the condition $\gamma_i = \sum_{k} g_{ik} \ll E_{M_i}$.
Similarly, the coupling between $M$ and $S$ fulfills the condition $g \ll E_{M_2}$, which ensures the validity of the weak-coupling limit~\cite{R6,R7}.

\subsection{Quantum Thermal Machine Modeling}
\label{sec:Quantum Thermal Machine Cycles}

The QATM consists of two qubits, $M_1$ and $M_2$, each in thermal equilibrium with its corresponding reservoir, $R_1$ and $R_2$, at temperatures $T_{R_i}=T_{M_i}$ for $i=\{1,2\}$. Their thermal states follow the Boltzmann distribution (with $k_B=1$):
\begin{align}
	\hat{\rho}_{M_i}(0) &= \frac{1}{Z_{M_i}} e^{-\frac{\hat{H}_{M_i}}{T_{M_i}}}, \quad i=\{1,2\}, \nonumber\\
	\hat{\rho}_{M}(0) &= \hat{\rho}_{M_1}(0) \otimes \hat{\rho}_{M_2}(0), 
	\label{D_QATM}
\end{align}
where $Z_{M_i}=\mathrm{Tr}_{M_i}\{e^{-\frac{\hat{H}_{M_i}}{T_{M_i}}}\}$ is the partition function of qubit $M_i$, and $\hat{\rho}_{M}(0)$ denotes the initial state of the QATM. The reservoirs $R_1$ and $R_2$ correspond to the cold and hot reservoirs, respectively, with $T_{M_1}\leq T_{M_2}$, while the energy spacings satisfy $E_{M_2}>E_{M_1}$. 

It is worth mentioning that during the dynamical evolution, the external system qubits are generally out of equilibrium with the QATM qubits. Then, the QATM qubits evolve toward thermal states that follow the Boltzmann distribution and eventually return to equilibrium with their respective reservoirs $R_1$ and $R_2$, approaching their corresponding Gibbs thermal states (see Fig.~\ref{TEMPVAR})\cite{R8}.

Since the Hamiltonians $\hat{H}_{M}$ and $\hat{H}_{S}$ are time-independent, the model allows only heat exchange between the subsystems and their reservoirs (see Appendix \ref{Autonomous Operation}). Accordingly, the heat fluxes are
\begin{align}
	Q_{MS}(t) &= \sum_{i=1}^{2} Q_{M_i}(t) + Q_{S}(t), \\
	Q_{M_i}(t) &= \mathrm{Tr}\{\hat{H}_{M_i} \, \Delta\hat{\rho}_{M_i}(t)\}, \nonumber \\
	Q_{S}(t) &= \mathrm{Tr}\{\hat{H}_{S} \, \Delta\hat{\rho}_{S}(t)\}, \nonumber
\end{align}
where $Q_{M_i}(t)$ is the heat exchanged between $M_i$ and $R_i$, and $Q_{S}(t)$ is the heat exchanged between the external system $S$ and the QATM $M$.  

The interactions among $R_1$, $R_2$, the QATM qubits, and the external system qubits $S_1$ and $S_2$, together with the temperature difference $T_{M_1} < T_{M_2}$ and the Hilbert space structure of each subsystem, give rise to two thermodynamic cycles on the QATM, denoted as Cycle A and Cycle B.  

\begin{figure*}[ht!]
	\subfloat[Cycle A\label{C_B}]{
		\includegraphics[width=1.01\columnwidth]{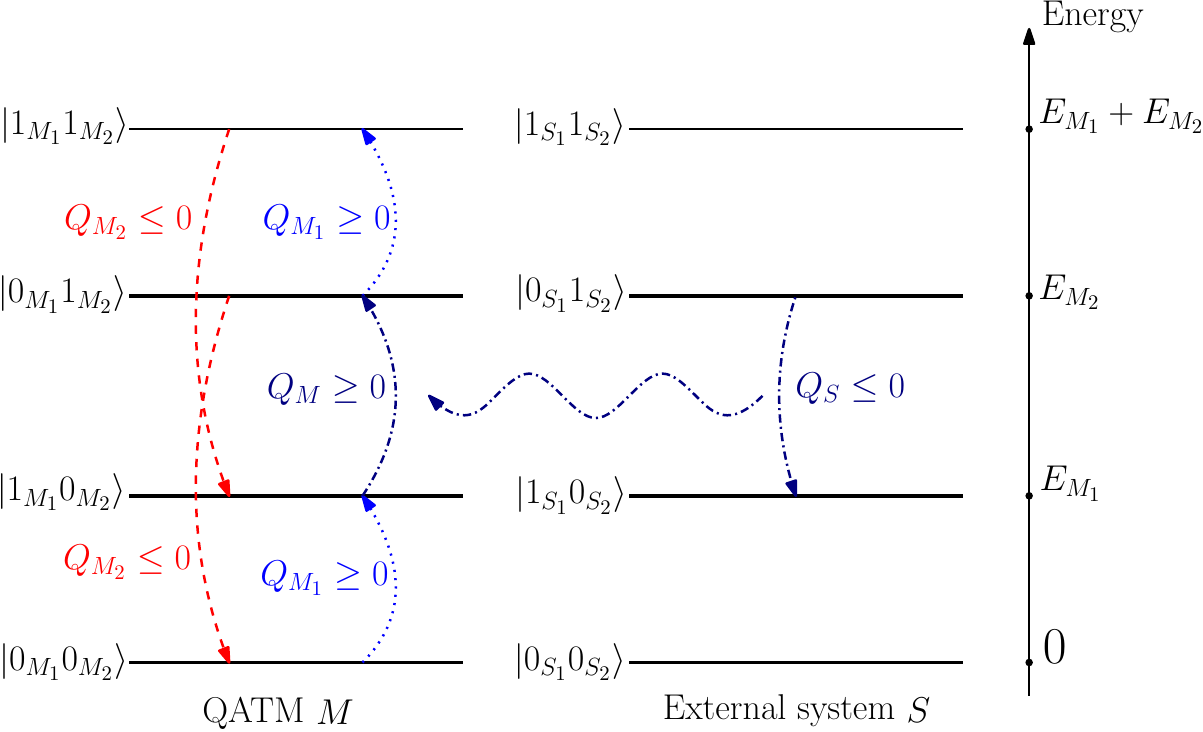}
	} \hfill
	\subfloat[Cycle B \label{C_A}]{
		\includegraphics[width=1.01\columnwidth]{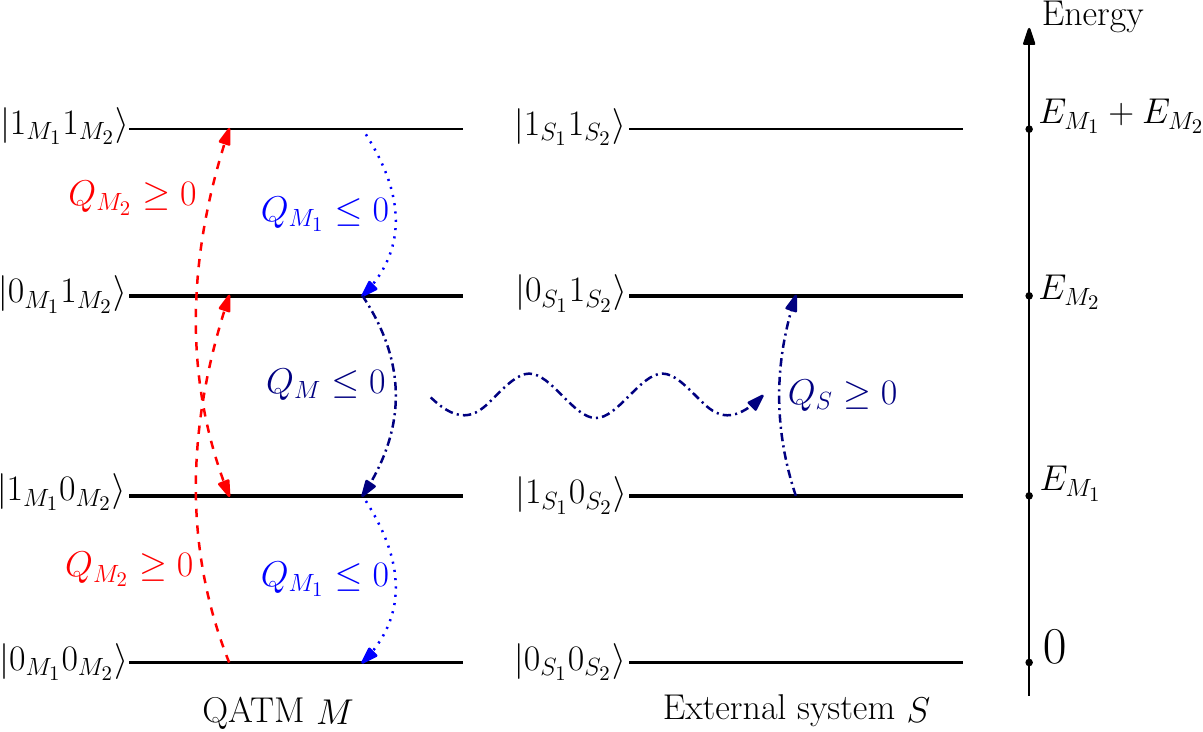}    
	}
	\caption{(a) Schematic representation of QATM Cycle A: heating of qubit $M_1$ and cooling of qubit $M_2$ with heat transfer from the external system to the machine.  
		(b) Cycle B: heating of qubit $M_2$ and cooling of qubit $M_1$ with heat transfer from the machine to the external system.  
		Arrows indicate the direction of heat flow between the qubits, thermal reservoirs, and the external system.}
	\label{C_AB}    
\end{figure*}

\textbf{Cycle A}, illustrated in Fig.~\ref{C_B}, involves the transitions $\ket{0_{M_1}0_{M_2}} \to \ket{1_{M_1}0_{M_2}}$ and $\ket{0_{M_1}1_{M_2}} \to \ket{1_{M_1}1_{M_2}}$, corresponding to heating of $M_1$ ($Q_{M_1}(t) \geq 0$). The transition $\ket{1_{M_1}0_{M_2}} \to \ket{0_{M_1}1_{M_2}}$ occurs, in which the machine absorbs heat from $S$ ($Q_M(t) \geq 0$), resulting in cooling of $S$ ($Q_S(t) \leq 0$). Finally, the transitions $\ket{0_{M_1}1_{M_2}} \to \ket{0_{M_1}0_{M_2}}$ and $\ket{1_{M_1}1_{M_2}} \to \ket{1_{M_1}0_{M_2}}$ correspond to cooling of $M_2$ ($Q_{M_2}(t) \leq 0$).  

\textbf{Cycle B}, illustrated in Fig.~\ref{C_A}, involves the transitions $\ket{1_{M_1}0_{M_2}} \to \ket{0_{M_1}0_{M_2}}$ and $\ket{1_{M_1}1_{M_2}} \to \ket{0_{M_1}1_{M_2}}$, corresponding to cooling of $M_1$ ($Q_{M_1}(t) \leq 0$). The transition $\ket{0_{M_1}1_{M_2}} \to \ket{1_{M_1}0_{M_2}}$ occurs, in which the machine releases heat to $S$ ($Q_M(t) \leq 0$), resulting in heating of $S$ ($Q_S(t) \geq 0$). Finally, the transitions $\ket{0_{M_1}0_{M_2}} \to \ket{0_{M_1}1_{M_2}}$ and $\ket{1_{M_1}0_{M_2}} \to \ket{1_{M_1}1_{M_2}}$ correspond to heating of $M_2$ ($Q_{M_2}(t) \geq 0$).  

The concept of a cycle follows the standard definition in quantum autonomous thermal machines \cite{INTRO9A,INTRO9B,INTRO9C,INTRO9D,INTRO10}. Unlike traditional stroke-based machines, where a cycle is defined by the evolution of external parameters such as bath temperatures \cite{INTRO1,INTRO2,INTRO3}, cycles A and B here refer to specific sequences of states traversed by the QATM system, consistent with the quantum autonomous framework.  

%%%%%%%%%%%%%%%%%%%%%%%%%%%FIG_1 %%%%%%%%%%%%%%%%%%%%%%%%%%%

The heat exchanged by the QATM is treated as a single body (see Appendix \ref{Virtual temperature})
\begin{align}
	Q_M(t) = E_M \big(P_M^E(t) - P_M^E(0)\big).
\end{align}

\begin{itemize}
	\item \textbf{Cycle A:} Heat flows from the external system to the QATM ($Q_M(t) \geq 0$), implying an increase in the QATM's excited-state population ($P_M^E(t) \geq P_M^E(0)$). Using Eqs.~\eqref{Popm} and \eqref{Virt}, the virtual temperature satisfies $T_M \leq 0$, leading to the constraint
	\begin{align}
		T_{M_1} \leq \frac{E_{M_1}}{E_{M_2}} T_{M_2}. \label{TconsitionA}
	\end{align}
	\item \textbf{Cycle B:} Heat flows from the QATM to the external system ($Q_M(t) \leq 0$), implying a decrease in the excited-state population ($P_M^E(t) \leq P_M^E(0)$). The virtual temperature satisfies $T_M \geq 0$, giving
	\begin{align}
		T_{M_1} \geq \frac{E_{M_1}}{E_{M_2}} T_{M_2}. \label{TconsitionB}
	\end{align}
\end{itemize}

For simplicity, we set $E_{M_i} = E_{S_i}$ for $i = \{1,2\}$, and the transition between Cycles A and B occurs at
\[
T_{M_1} = \frac{E_{M_1}}{E_{M_2}} T_{M_2}.
\]

In the simulations, we use $T_{M_1} = 0.1 T_{M_2}$ for Cycle A and $T_{M_1} = 0.8 T_{M_2}$ for Cycle B, with $T_{M_2} = E_{M_2} = 10$, and couplings $g = 0.03E_{M_2}, 0.05E_{M_2}, 0.07E_{M_2}, 0.09E_{M_2}$. Dissipation rates are $\gamma_i = 0.01E_{M_i}$ for $i = \{1,2\}$.  

These parameters also map the theoretical model to experimental realizations. For superconducting qubits \cite{INTRO31,INTRO32,INTRO33}, energy spacings lie between 4~GHz and 10~GHz. In natural units ($\hbar = 1$), $E_{M_2} = 10$ corresponds to the GHz range, while $g$ ranges from 10~MHz to 100~MHz and the time scale spans 20--200~$\mu$s, demonstrating the model's experimental feasibility.  

\section{Dynamics of Quantum Thermodynamic Quantities: Heat and Temperature}\label{sec:Dynamics of Quantum Thermodynamic Quantities: Heat and Temperature}
In this section, we numerically investigate the constraints provided in Eq.~\ref{TconsitionA} and Eq.~\ref{TconsitionB} on the heat dynamics of each subsystem and on the temperature dynamics. As the initial state of the QATM $M$, we use the density matrix given in Eq.~\ref{D_QATM}. For the external system $S$, throughout the remainder of the paper, we consider the superposition state $\ket{\psi(0)}_{S_1} = \frac{1}{\sqrt{2}}(\ket{0}_{S_1} + \ket{1}_{S_1})$ for qubit $S_1$, and the superposition state $\ket{\psi(0)}_{S_2} = \frac{1}{\sqrt{2}}(\ket{0}_{S_2} - \ket{1}_{S_2})$ for qubit $S_2$. The corresponding initial density matrix is given by $\hat{\rho}_{S_i}(0) = \ket{\psi(0)}_{S_i}\bra{\psi(0)}_{S_i}$ for $i = \{1, 2\}$.

\subsection{Heat dynamics}
\label{sec:Heat dynamics}
We analyze the heat dynamics of the QATM qubits $M_{1}$ and $M_{2}$, denoted by $Q_{M_1}(t)$ and $Q_{M_2}(t)$, respectively, as well as those of the external system qubits $S_1$ and $S_2$, denoted by $Q_{S_1}(t)$ and $Q_{S_2}(t)$, respectively. We set the coupling strength between the QATM $M$ and the external system $S$ to $g=0.09$, and fix the temperature of qubit $M_2$ to $T_{M_2}=1$. The heats are then analyzed as functions of time and the temperature $T_{M_{1}}$, which is varied in the interval $T_{M_1} \in [0.1T_{M_2},\, T_{M_2}]$, and represented using contour plots.

%%%%%%%%%%%%%%%%%%%%%%%%%%%FIG_2%%%%%%%%%%%%%%%%%%%%%%%%%%%%%%%%%%%%%%%%%%%%%%
\begin{figure*}[ht!]
	\subfloat[\label{Qm1T1}]{
		\includegraphics[width=1
		\columnwidth]{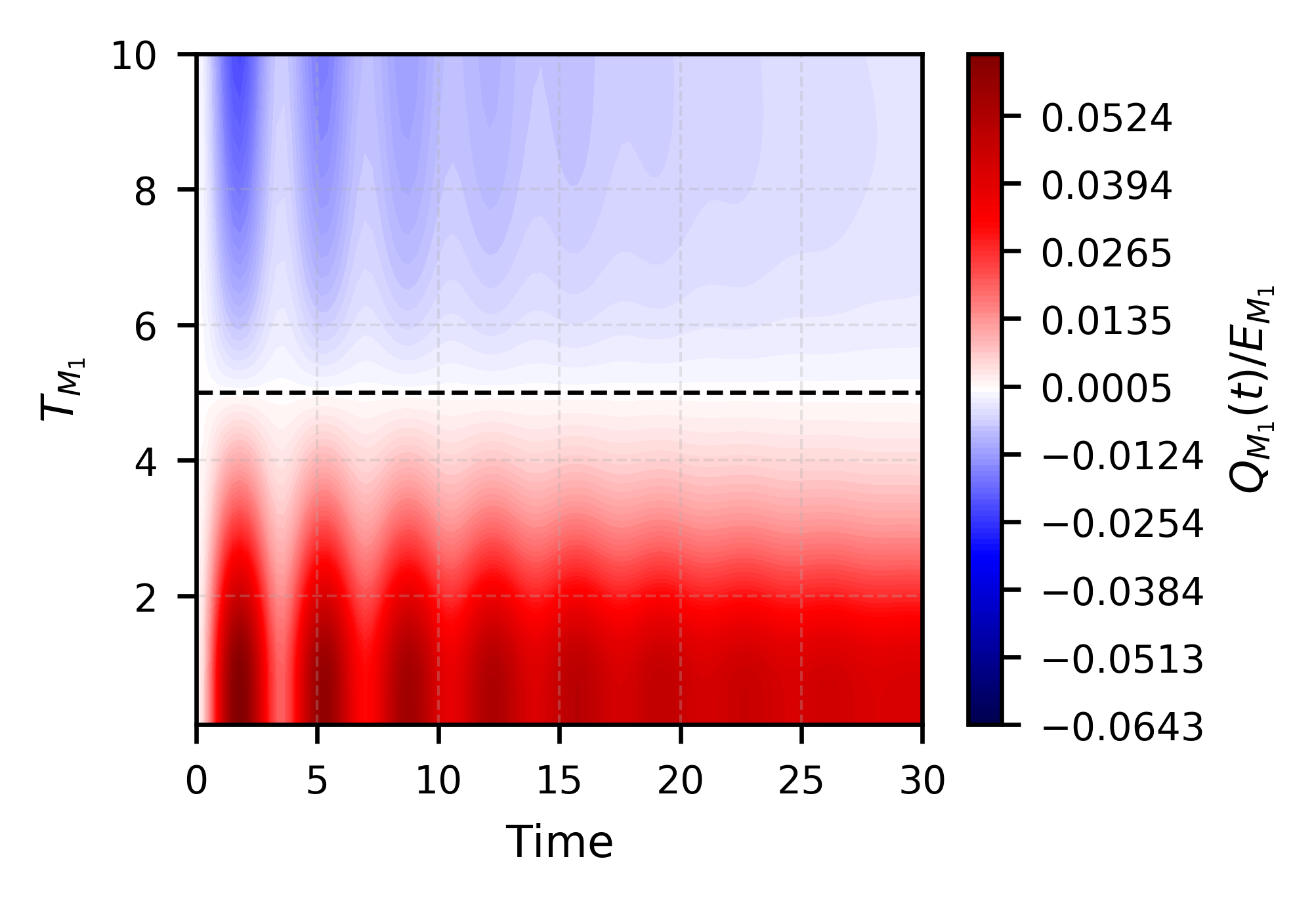}
	} \hfill
	\subfloat[\label{Qm2T1}]{
		\includegraphics[width=1
		\columnwidth]{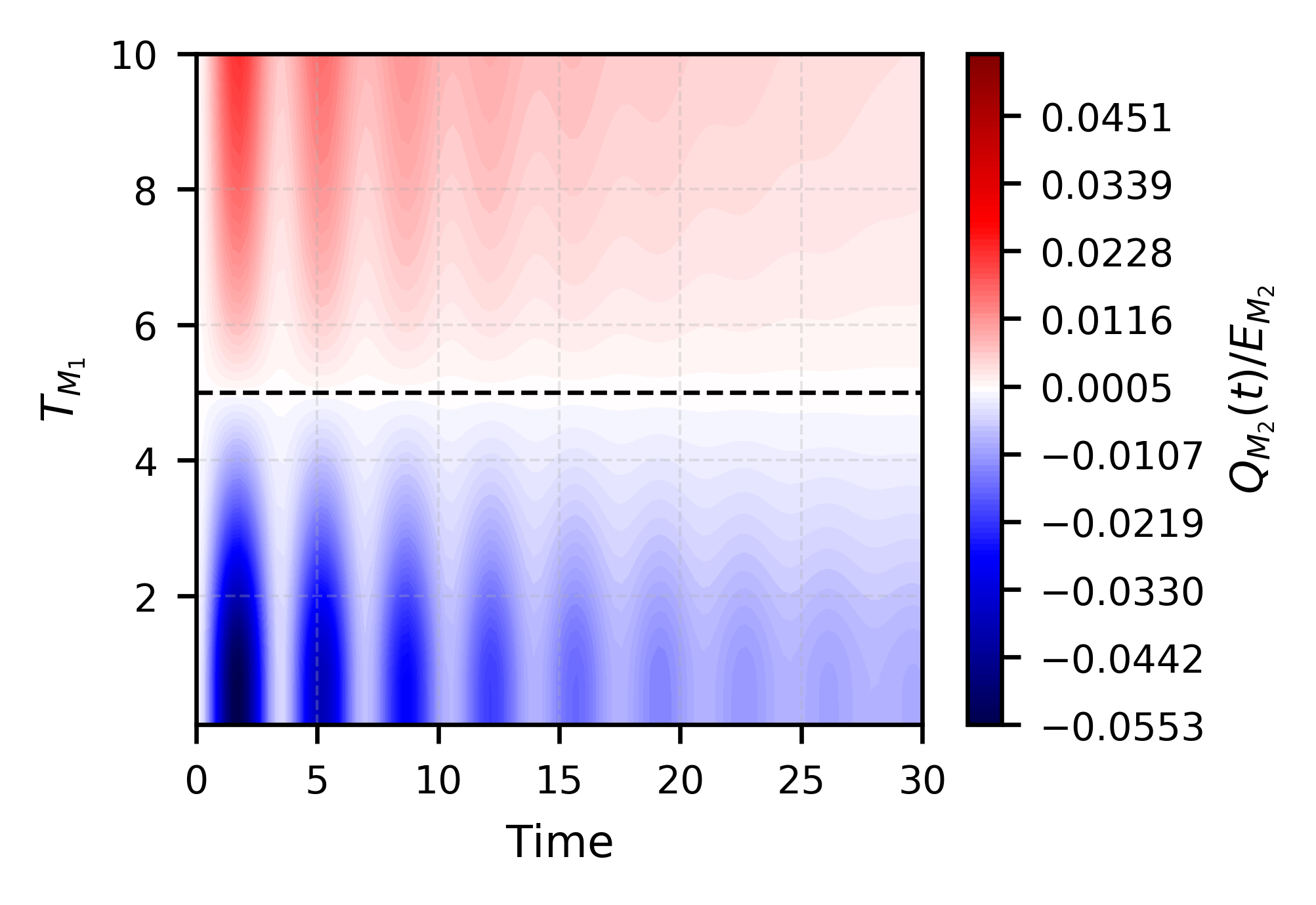}	
	}\hfill
	\subfloat[\label{Qs1T1}]{
		\includegraphics[width=1
		\columnwidth]{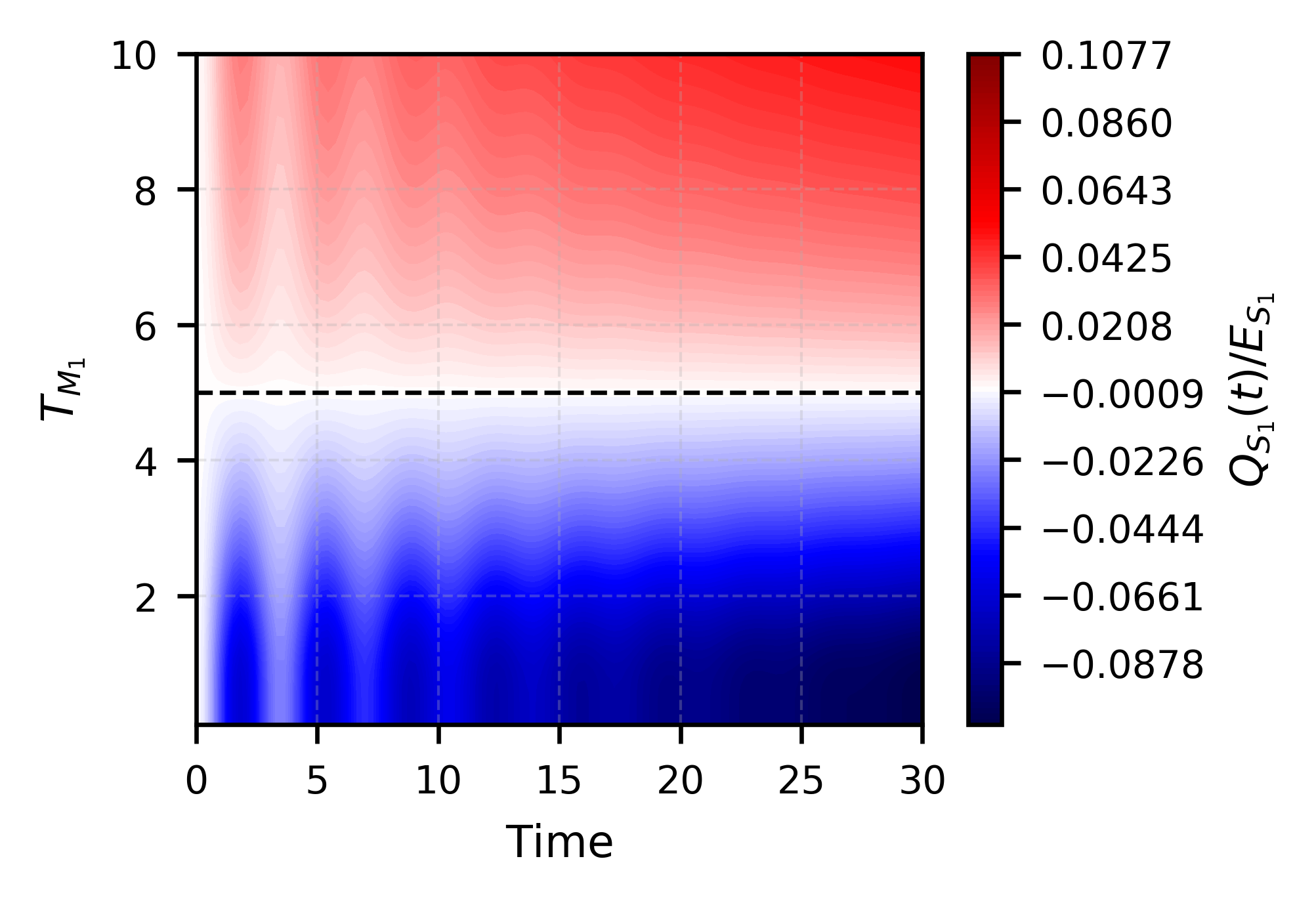}
	} \hfill
	\subfloat[\label{Qs2T1}]{
		\includegraphics[width=1
		\columnwidth]{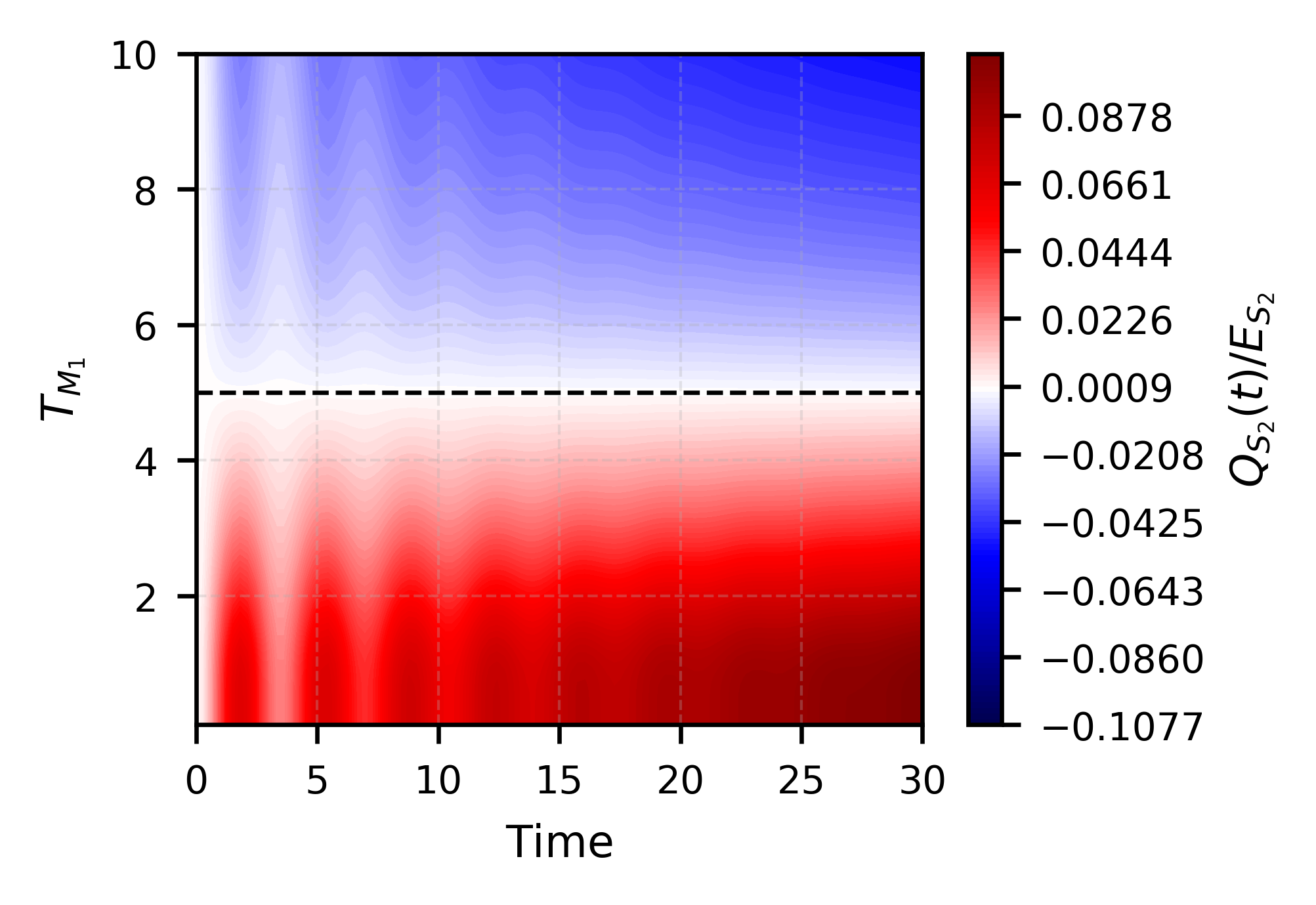}	
	}
	\caption{Heat dynamics for each qubit in the QATM and in the external system.  
		(a,b) Heat exchanged $Q_{M_1}(t)$ and $Q_{M_2}(t)$ for the QATM qubits.  
		(c,d) Heat exchanged $Q_{S_1}(t)$ and $Q_{S_2}(t)$ for the external system qubits.  
		The two regions separated by the black dashed line correspond to the conditions of cycles A and B according to the ratio $T_{M_1}/T_{M_2}=0.5$, we set $T_{M_2}=E_{M_2}$, and $g=0.08 E_{M_2}$.}
	\label{HT}	
\end{figure*}
%%%%%%%%%%%%%%%%%%%%%%%%%%%%%%%%%%%%%%%%%%%%%%%%%%%%%%%%%%%%%%%%%%%%%%%%%%%%%%%%%%%%%%%%%%%%%%%

The variations of the heat for the QATM qubits, $Q_{M_1}(t)$ and $Q_{M_2}(t)$, are shown in Fig.~\ref{Qm1T1} and Fig.~\ref{Qm2T1}, respectively. We observe that when $T_{M_1} \leq 0.5T_{M_2}$, the heat satisfies $Q_{M_1}(t) \geq 0$ and $Q_{M_2}(t) \leq 0$. In contrast, when $T_{M_1} \geq 0.5T_{M_2}$, the heat satisfies $Q_{M_1}(t) \leq 0$ and $Q_{M_2}(t) \geq 0$. Physically, this behavior corresponds to the transition conditions between cycles A and B described in Eq.~\ref{TconsitionA} and Eq.~\ref{TconsitionB}.

For the external system qubits $S_1$ and $S_2$, the heat variations are shown in Fig.~\ref{Qs1T1} and Fig.~\ref{Qs2T1}, respectively. When $T_{M_1} \leq 0.5T_{M_2}$ (cycle A), the heats satisfy $Q_{S_1}(t) \leq 0$ and $Q_{S_2}(t) \geq 0$. Conversely, when $T_{M_1} \geq 0.5T_{M_2}$ (cycle B), the heats satisfy $Q_{S_1}(t) \geq 0$ and $Q_{S_2}(t) \leq 0$. 

In connection with the variation of the QATM qubits heat $Q_{M_i}(t)$ for $i \in \{1,2\}$, and under the resonance condition between qubits $M_i$ and $S_i$ ($E_{M_i} = E_{S_i}$), the heat is partially exchanged as $Q_{S_i}(t) \rightarrow Q_{M_i}(t)$ for cycle A, and $Q_{M_i}(t) \rightarrow Q_{S_i}(t)$ for cycle B.

The non-monotonic behavior of the heat dynamics for $Q_{M_i}(t)$ and $Q_{S_i}(t)$ arises from memory effects (backflow of information between the subsystems) present in our theoretical model, as will be shown in the next subsection, Sec.~\ref{sec:Non-Markovianity and correlation exchanges between the QATM and the external system}.

The asymmetry observed in the dynamical quantities around zero reflects the intrinsic nonequilibrium nature of the autonomous thermal machine studied in this work. In particular, the system operates under a structured energy configuration satisfying $E_{M_2} - E_{M_1} = E_{S_2} - E_{S_1}$ with $E_{M_2} > E_{M_1}$ and $E_{S_2} > E_{S_1}$. Together with the resonance mismatch between the subsystems and their collective interaction with the environment, this leads to non-equivalent energy exchange channels and a natural breaking of symmetry between positive and negative contributions. This behavior is therefore a genuine physical feature of the nonequilibrium dynamics rather than a numerical artifact.

\subsection{Temperature dynamics}
\label{sec:Temperature dynamics}
To illustrate the dynamics of the temperatures of the QATM qubits $T_{M_i}$ for qubits $M_i$ with $i \in \{1,2\}$, as well as the virtual temperature of the QATM $T_M$, we analyze Eq.~\ref{Popm}. We obtain
\begin{align}
	T_{M_i}(t) &= \dfrac{E_{M_i}}{\ln\!\left(P_{M_i}^{G}(t)\right) - \ln\!\left(P_{M_i}^{E}(t)\right)}, \quad (i \in \{1,2\}), \nonumber\\
	T_{M}(t) &= \dfrac{E_{M}}{\ln\!\left(P_{M}^{G}(t)\right) - \ln\!\left(P_{M}^{E}(t)\right)}.
\end{align}

%%%%%%%%%%%%%%%%%%%%%%%%%%%%Fig_3%%%%%%%%%%%%%%%%%%%%%%%%%%%%%%%%%%%%%%%%%%%%%
\begin{figure*}[ht!]
	\subfloat[Cycle A ($T_{M_1}=0.1T_{M_2}$)\label{TEMPERATURESCYCLEA}]{
		\includegraphics[width=1
		\columnwidth]{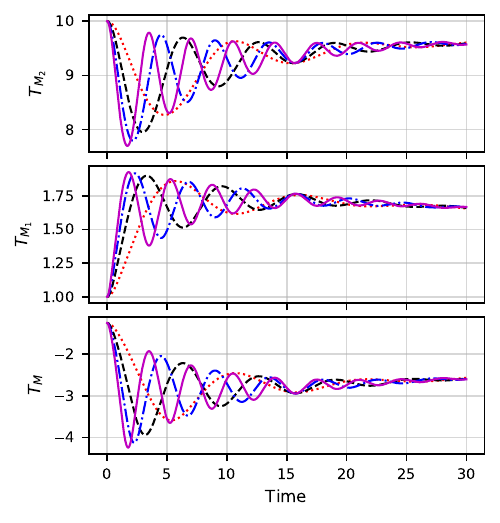}
	} \hfill
	\subfloat[Cycle B ($T_{M_1}=0.8T_{M_2}$)\label{TEMPERATURESCYCLEB}]{
		\includegraphics[width=1
		\columnwidth]{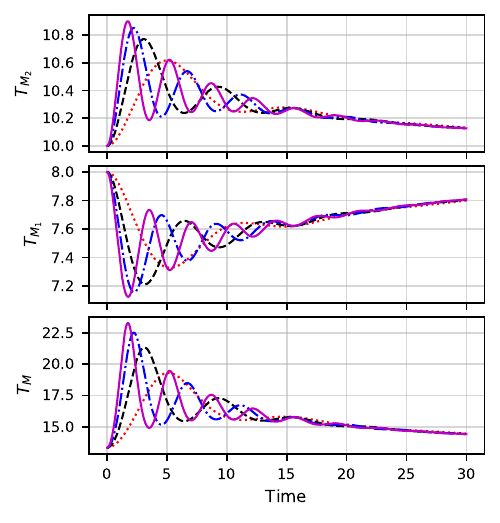}	
	}
	\caption{Temperature evolution of the QATM qubits over time.  
		(a) Cycle A: gradual increase of $T_{M_1}$ and decrease of $T_{M_2}$ with a negative virtual temperature $T_M < 0$.  
		(b) Cycle B: decrease of $T_{M_1}$ and increase of $T_{M_2}$ with a positive virtual temperature $T_M > 0$. The coupling strength is varied as $g=0.03,\,0.05,\,0.07,$ and $0.09$ (in units of $E_{M_2}$), represented by red (dotted), black (dashed), blue (dot--dashed), and magenta (solid) curves.}
	\label{TEMPVAR}	
\end{figure*}
%%%%%%%%%%%%%%%%%%%%%%%%%%%%%%%%%%%%%%%%%%%%%%%%%%%%%%%%%%%%%%%%%%%%%%%%%%%%%%%%%%%%%%%%%%%

For cycle A in Fig.~\ref{TEMPERATURESCYCLEA}, the initial conditions are $T_{M_1} = 0.1T_{M_2}$, $T_{M_2} = 1$, and $T_{M} \leq 0$. Over time, the temperature of qubit $M_1$ increases, while that of $M_2$ decreases, in agreement with the heat transfer shown in Fig.~\ref{Qm1T1} and Fig.~\ref{Qm2T1}, under the condition given in Eq.~\ref{TconsitionA}. Physically, during the transition of qubit $M_1$ from $\ket{0}_{M_1}$ to $\ket{1}_{M_1}$, it absorbs positive heat $Q_{M_1}(t) \geq 0$ from its reservoir $R_1$. In contrast, qubit $M_2$, during the transition from $\ket{1}_{M_2}$ to $\ket{0}_{M_2}$, releases heat $Q_{M_2}(t) \leq 0$ to its reservoir $R_2$. Consequently, the QATM $M$ absorbs heat $Q_{M} \geq 0$ from the external system $S$ at the virtual temperature $T_v$.

For cycle B in Fig.~\ref{TEMPERATURESCYCLEB}, the initial conditions are $T_{M_1} = 0.8T_{M_2}$, $T_{M_2} = 1$, and $T_{M} \geq 0$. Over time, the temperature of qubit $M_1$ decreases while exchanging positive heat $Q_{M_1}(t) \geq 0$ with its reservoir $R_1$ (cooling), whereas the temperature of qubit $M_2$ increases while exchanging negative heat $Q_{M_2}(t) \leq 0$ with its reservoir $R_2$ (heating), satisfying the condition in Eq.~\ref{TconsitionB}.

The non-monotonic evolution of the qubit temperatures with increasing coupling between the QATM $M$ and the external system $S$ originates from memory effects, which correspond to information exchange between the QATM and the external system. Note that the virtual temperature can take negative values; this is not unphysical, as it is associated with population inversion (as in lasers). The virtual temperature plays an important role in complex quantum thermodynamic systems, allowing one to distinguish between different types of thermal machines.

\subsection{Production entropy and production entropy rate}
\label{sec:Production entropy and production entropy rate}
The entropy production and the entropy production rate, denoted by $\Sigma(t)$ and $\dot{\Sigma}(t)$, respectively, are fundamental concepts in quantum thermodynamics. They quantify the irreversibility of an open quantum system over time and can also be interpreted as a measure of the information loss from the system to its environment. Mathematically, they are defined as
\begin{align}
	\Sigma (t) &= \Delta S(\hat{\rho}(t)) - \sum_{i=1}^{2} \frac{\Delta Q_{M_i}(t)}{T_{M_i}(t)}, \\
	\sigma(t) &= \frac{d}{dt}\Sigma (t),
\end{align}
where $S(\hat{\rho}(t)) = -\mathrm{Tr}\{\hat{\rho}(t)\log_2[\hat{\rho}(t)]\}$ denotes the von Neumann entropy of the total density matrix of our theoretical model. We define $\Delta S(\hat{\rho}(t)) = S(\hat{\rho}(t)) - S(\hat{\rho}(0))$ and $\Delta Q_{M_i}(t) = Q_{M_i}(t) - Q_{M_i}(0) = Q_{M_i}(t)$ for the thermal machine qubit $M_i$, with $i \in \{1, 2\}$.

%%%%%%%%%%%%%%%%%%%%%%%%%%%%Fig_4%%%%%%%%%%%%%%%%%%%%%%%%%%%%%%%%%%%%%%%%%%%%%
\begin{figure*}[ht!]
	\subfloat[Cycle A ($T_{M_1}=0.1T_{M_2}$)\label{ENTROPYPRODUCTIONCYCLEA}]{
		\includegraphics[width=1
		\columnwidth]{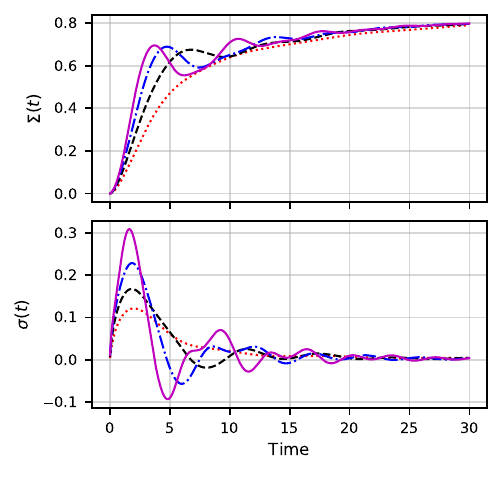}
	} \hfill
	\subfloat[Cycle B ($T_{M_1}=0.8T_{M_2}$)\label{ENTROPYPRODUCTIONCYCLEB}]{
		\includegraphics[width=1
		\columnwidth]{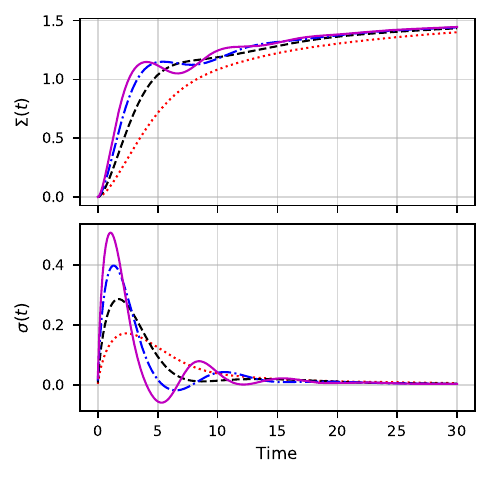}	
	}
	\caption{Entropy production $\Sigma(t)$ and entropy production rate $\sigma(t)$ over time.  
		(a) Cycle A: smaller entropy production with more frequent and stronger negative rates, indicating stronger non-Markovianity.  
		(b) Cycle B: larger entropy production with less frequent negative rates, indicating weaker non-Markovianity. The coupling strength is varied as $g=0.03,\,0.05,\,0.07,$ and $0.09$ (in units of $E_{M_2}$), represented by red (dotted), black (dashed), blue (dot--dashed), and magenta (solid) curves.}
	\label{EP}	
\end{figure*}
%%%%%%%%%%%%%%%%%%%%%%%%%%%%%%%%%%%%%%%%%%%%%%%%%%%%%%%%%%%%%%%%%%%%%%%%%%%%%%%%%%%%%%%%%%%
In Fig.~\ref{EP}, we analyze the entropy production and the entropy production rate as functions of time and of the coupling strength between the QATM $M$ and the external system $S$, for cycles A and B shown in Fig.~\ref{ENTROPYPRODUCTIONCYCLEA} and Fig.~\ref{ENTROPYPRODUCTIONCYCLEB}, respectively. For both cycles A and B, the entropy production $\Sigma(t)$ remains positive over time. Moreover, the entropy production in cycle A is smaller than that in cycle B throughout the evolution. Physically, this indicates that the irreversibility of the total system increases over time and that the decoherence effects induced by the reservoirs $R_1$ and $R_2$ are more pronounced in cycle B than in cycle A. This means that the loss of information from the QATM and the external system is more significant in cycle B than in cycle A.

For the entropy production rate $\sigma(t)$, we observe a non-monotonic evolution over time for both cycles A and B. In addition, $\sigma(t)$ takes negative values during the evolution. Moreover, in cycle A, $\sigma(t)$ assumes negative values more frequently and with larger magnitude than in cycle B. Physically, negative values of the entropy production rate, i.e., $\sigma(t) < 0$, indicate the presence of non-Markovian dynamics during the evolution of the total system $MS$. Therefore, as mentioned previously, the fact that $\sigma(t)$ reaches more negative values in cycle A suggests that the dynamics are more non-Markovian in cycle A than in cycle B. Furthermore, increasing the coupling strength $g$ between the QATM $M$ and the external system $S$ enhances the non-Markovian behavior, since larger values of $g$ lead to more pronounced negative values of $\sigma(t)$ compared to smaller values of $g$.

In the next section, we examine the effect of the QATM cycle on the dynamics of the external system qubits, identify the origin of the non-Markovianity, and discuss the implications of these results for entanglement generation in the external system $S$.
\section{Impact of Memory Effects on Entanglement Generation in QATM Cycles for External System Qubits}
\label{sec:Impact of Memory Effects on Entanglement Generation in QATM Cycles for External System Qubits}
As discussed previously, signatures of non-Markovianity are observed through oscillations in the heat transfer, the temperature evolution of the QATM qubits, and the negative values of the entropy production rate. In this section, we investigate the origin of the non-Markovianity and analyze the effect of QATM cycles A and B on entanglement generation and the role of coherence in the external system $S$.

\subsection{Non-Markovianity and correlation exchanges between the QATM and the external system}
\label{sec:Non-Markovianity and correlation exchanges between the QATM and the external system}

In our previous work~\cite{INTRO10}, we showed that the QATMs can induce a backflow of information (non-Markovianity) in the external system through the exchange of correlations. Similarly, Popovic \textit{et al.}~\cite{INTRO25} demonstrated that non-Markovianity can arise from correlation exchanges between subsystems. In our scenario, we quantify non-Markovianity using the measure proposed by Breuer \textit{et al.}~\cite{INTRO26}, denoted by $\mathcal{N}_{m}(t)$, which is based on the distinguishability between two quantum states $\hat{\rho}_{m}^{\alpha}(t)$ and $\hat{\rho}_{m}^{\beta}(t)$ for $m \in \{M, S\}$. Mathematically, it is defined as~\cite{INTRO26}
	\begin{equation}
		\mathcal{N}_{m}(t) = \max_{\hat{\rho}_{m}^{\alpha\beta}(0)} \int_{\vartheta_{m}(t) > 0} dt\, \vartheta_{m}(t),
	\end{equation}
	where $\hat{\rho}_{m}^{\alpha\beta}(0) = \hat{\rho}_{m}^{\alpha}(0) - \hat{\rho}_{m}^{\beta}(0)$ and $\vartheta_{m}(t) = \frac{d}{dt} \left\| \hat{\rho}_{m}^{\alpha\beta}(t) \right\|$ is the time derivative of the trace distance between $\hat{\rho}_{m}^{\alpha}(t)$ and $\hat{\rho}_{m}^{\beta}(t)$.

However, we quantify the amount of quantum correlations between the QATM $M$ and the external system $S$ using the quantum mutual information, namely $\mathcal{I}_{M-S}(t)$, which is defined as~\cite{INTRO27}
	\begin{equation}
		\mathcal{I}_{M-S}(t) = S\big(\hat{\rho}_{M}(t)\big) + S\big(\hat{\rho}_{S}(t)\big) - S\big(\hat{\rho}(t)\big),
	\end{equation}
	where $S(\hat{\rho}_{m}(t))$ is the von-Neumann entropy of the subsystem $m \in \{M,S\}$. Moreover, for the initial state of the QATM $M$, we set $\hat{\rho}_{M}^{\alpha}(0) = \hat{\rho}_{M}^{\beta}(0) = \hat{\rho}_{M}(0)$ as in Eq.~\ref{D_QATM}. For the external system $S$, we take $\hat{\rho}_{S}^{\alpha}(0) = \hat{\rho}_{S}(0)$ and $\hat{\rho}_{S}^{\beta}(0) = \bar{\rho}_{S}(0)$, where $\hat{\rho}_{S}(0)$ is the state used in Sec.~\ref{sec:Dynamics of Quantum Thermodynamic Quantities: Heat and Temperature}. The state $\bar{\rho}_{S}(0)$ corresponds to the fully dephased (incoherent) state of the external system $S$ in the computational basis,
	\begin{equation}
		\bar{\rho}_{S}(0) = \sum_{n_S} \braket{n_S | \hat{\rho}_{S}(0) | n_S} \, \ket{n_S}\bra{n_S}.
\end{equation}

%%%%%%%%%%%%%%%%%%%%%%%%%%%%Fig_5%%%%%%%%%%%%%%%%%%%%%%%%%%%%%%%%%%%%%%%%%%%%%
\begin{figure*}[ht!]
	\subfloat[Cycle A ($T_{M_1}=0.1T_{M_2}$)\label{NONMARKOVIANITYCYCLEA}]{
		\includegraphics[width=1\columnwidth]{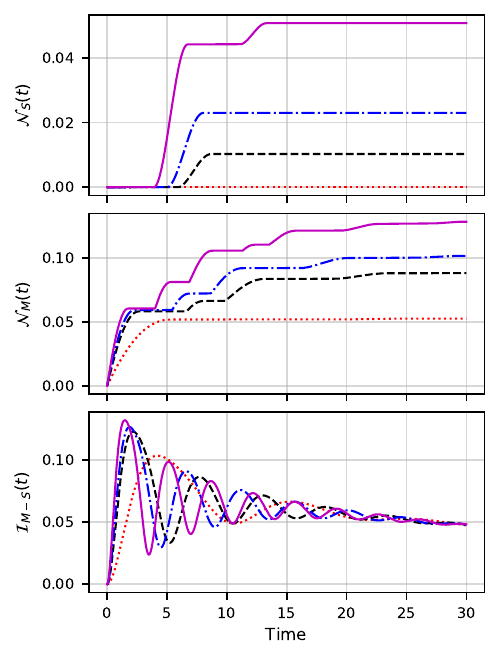}
	} \hfill
	\subfloat[Cycle B ($T_{M_1}=0.8T_{M_2}$)\label{NONMARKOVIANITYCYCLEB}]{
		\includegraphics[width=1\columnwidth]{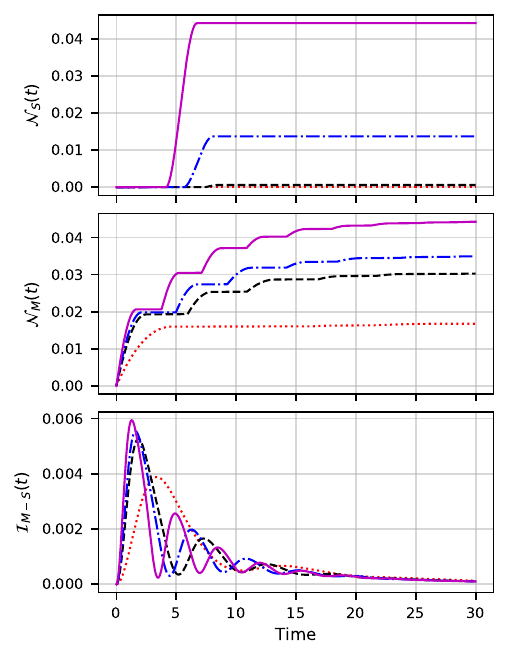}
	}
	\caption{Non-Markovianity measures and mutual information between the QATM and the external system.  
		(a) Cycle A: higher non-Markovianity in the subsystems and stronger mutual information $\mathcal{I}_{M-S}(t)$ as the coupling $g$ increases.  
		(b) Cycle B: lower non-Markovianity and mutual information under the same coupling conditions. 
		The coupling strength is varied as $g=0.03,\,0.05,\,0.07,$ and $0.09$ (in units of $E_{M_2}$), represented by red (dotted), black (dashed), blue (dot--dashed), and magenta (solid) curves.}
	\label{NM}
\end{figure*}
%%%%%%%%%%%%%%%%%%%%%%%%%%%%%%%%%%%%%%%%%%%%%%%%%%%%%%%%%%%%%%%%%%%%%%%%%%%%%%%%%%%%%%%%%%%

In Fig.~\ref{NM}, we analyze the non-Markovianity measures $\mathcal{N}_{M}(t)$ and $\mathcal{N}_{S}(t)$ for the QATM $M$ and the external system $S$, respectively, as well as the mutual information $\mathcal{I}_{M-S}(t)$ between $M$ and $S$, versus time and coupling strength $g$. These results are shown for cycles $A$ and $B$ in Figs.~\ref{NONMARKOVIANITYCYCLEA} and~\ref{NONMARKOVIANITYCYCLEB}, respectively.
For both cycles, increasing the coupling strength leads to an increase in the amplitude of non-Markovianity in the subsystems $M$ and $S$, accompanied by a growth in correlations between them, which is particularly pronounced at short times. For robust values of time, the correlations decrease non-monotonically due to the decoherence effects, while the observed non-monotonic behavior reflects information backflow between the subsystems. Moreover, both non-Markovianity and correlations are dominant in cycle $A$ than in cycle $B$, confirming that negative values of the entropy production rate serve as indicators of memory effects in our model. The larger correlations observed in cycle $A$ also indicate that this cycle preserves information between the subsystems more effectively than cycle $B$.

The memory effects in our model arise from the exchange of correlations between the QATM $M$ and the external system $S$ over time, where these correlations partially preserve information between the subsystems. In this regard, the QATM can be interpreted as a quantum-structured non-Markovian reservoir interacting with the reservoirs $R_1$ and $R_2$, effectively acting as a filter for the decoherence effects of the reservoirs on the external system $S$.

Note that the master equation in Eq.~\ref{MASTER_EQUATION} describes Markovian dynamics for the following total density matrix: 
	\begin{equation}
		\hat{\rho}_{SM}(t) = \hat{\rho}_{M}(t) \otimes \hat{\rho}_{S}(t) + \hat{\rho}_{MS}^{\mathrm{corr}}(t),
	\end{equation}
where $\hat{\rho}_{MS}^{\mathrm{corr}}(t)$ accounts for correlations between $M$ and $S$ governed by the interaction Hamiltonian $\hat{H}_{MS}$, which is quantified by means of $\mathcal{I}_{M-S}(t)$. When investigating the partial trace over $S$ or $M$, the resulting reduced dynamics is not necessarily Markovian. The observed non-Markovian behavior in the subsystems arises from the exchange of correlations between $S$ and $M$ over time. In this inspiration, non-Markovianity refers specifically to the reduced dynamics of the subsystems, particularly the external system $S$. Although the full $MS$ evolution is of Lindblad form and therefore Markovian, tracing out part of the composite system generally leads to effective reduced dynamics that are not necessarily completely positive (CP) divisible map. Hence, memory effects can emerge at the level of the subsystems.
\subsection{Entanglement generation }\label{sec:Entangelment generaation}
To illustrate the effect of QATM $M$ cycles A and B on entanglement generation in the external system $S$ over time, we quantify the entanglement using the concurrence~\cite{INTRO28}, denoted by $\mathcal{C}_{S_1 - S_2}(t)$, and the correlation between the external qubits $S_1$ and $S_2$ using the mutual information~\cite{INTRO27}, denoted by $\mathcal{I}_{S_1 - S_2}(t)$. These quantities are mathematically defined as
\begin{align}
	\mathcal{C}_{S_1 - S_2}(t) &= \max\big[0, \lambda_1(t) - \sum_{i\neq 1}^{4} \lambda_i(t) \big], \\
	\mathcal{I}_{S_1 - S_2}(t) &= S(\hat{\rho}_{S_1}(t)) + S(\hat{\rho}_{S_2}(t)) - S(\hat{\rho}_{S}(t)),
\end{align}
where $\lambda_1(t) \geq \lambda_2(t) \geq \lambda_3(t) \geq \lambda_4(t)$ are the eigenvalues of the spin-flipped density matrix
\begin{equation}
	\hat{\rho}_{S}(t) (\sigma^y_{S_1} \otimes \sigma^y_{S_2}) \hat{\rho}_{S}^*(t) (\sigma^y_{S_1} \otimes \sigma^y_{S_2}),
\end{equation}
and $S(\hat{\rho}_{S_i}(t))$ is the von Neumann entropy of qubit $S_i$ for $i \in \{1,2\}$.
%%%%%%%%%%%%%%%%%%%%%%%%%%%%Fig_6%%%%%%%%%%%%%%%%%%%%%%%%%%%%%%%%%%%%%%%%%%%%%
\begin{figure*}[ht!]
	\subfloat[\label{CONCURANCETEMPERATURES}]{
		\includegraphics[width=1
		\columnwidth]{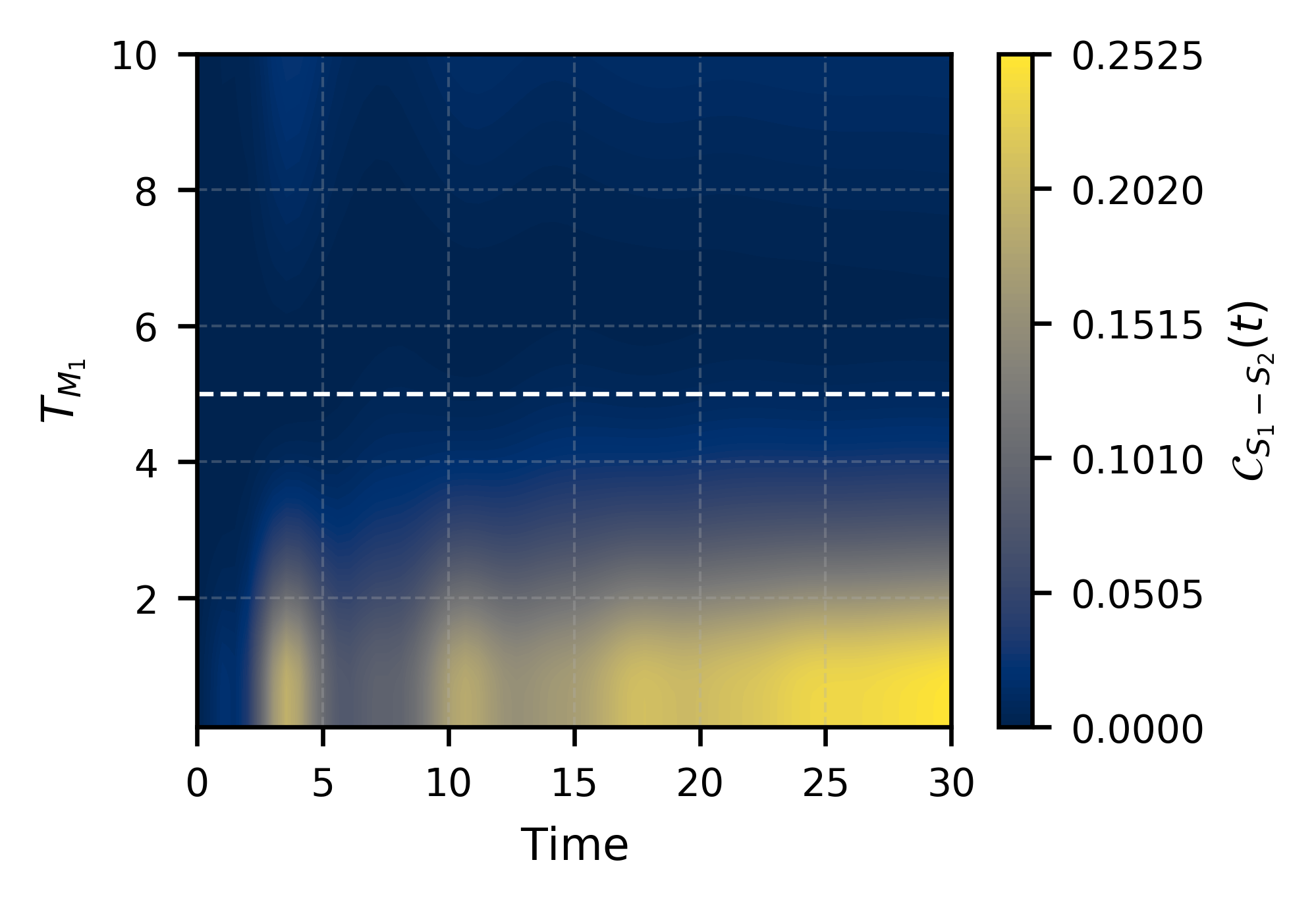}
	} \hfill
	\subfloat[\label{CORRELATION}]{
		\includegraphics[width=1
		\columnwidth]{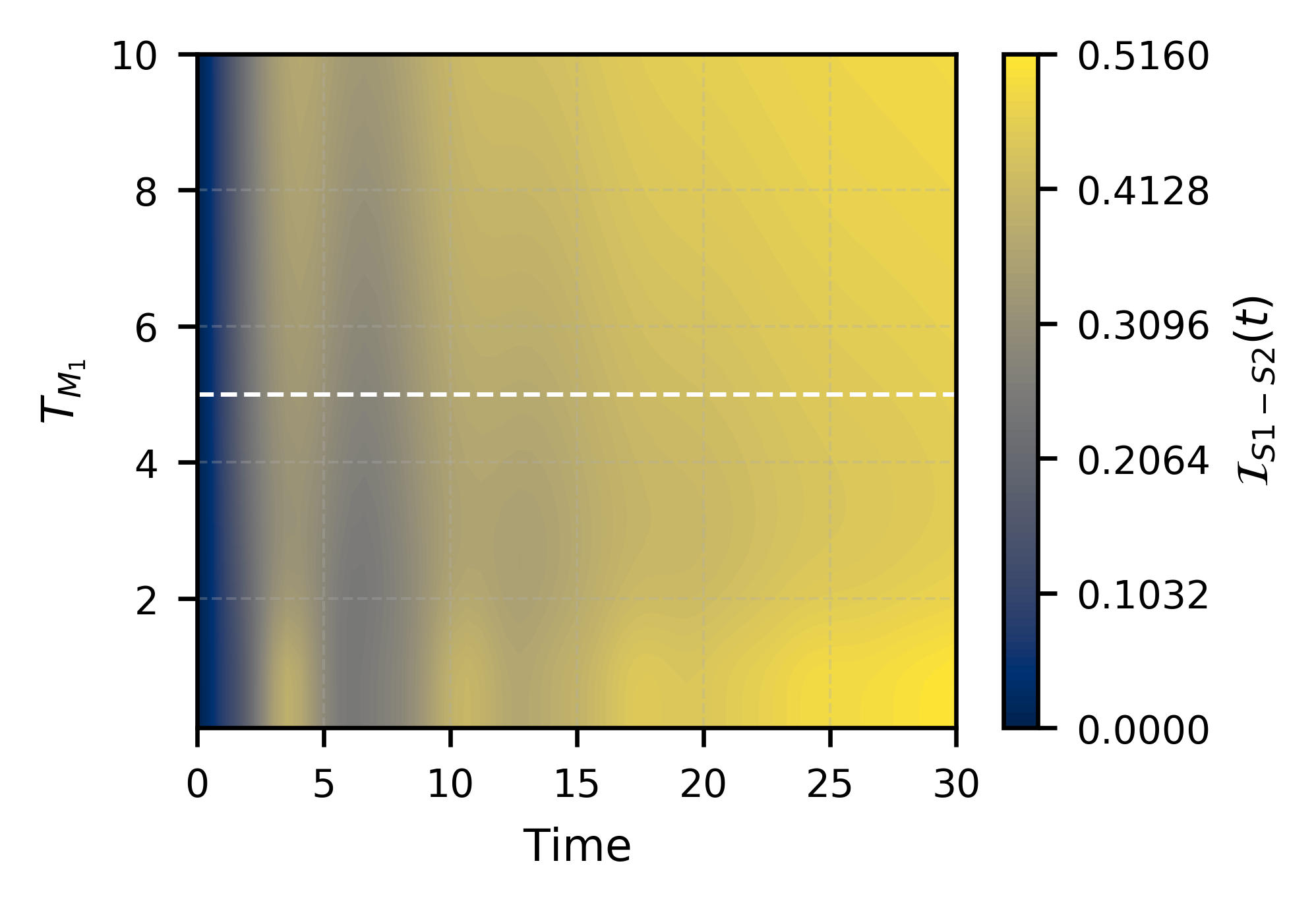}	
	}
	\hfill
	\subfloat[Cycle A ($T_{M_1}=0.1T_{M_2}$)\label{CORRELATIONCA}]{
		\includegraphics[width=1
		\columnwidth]{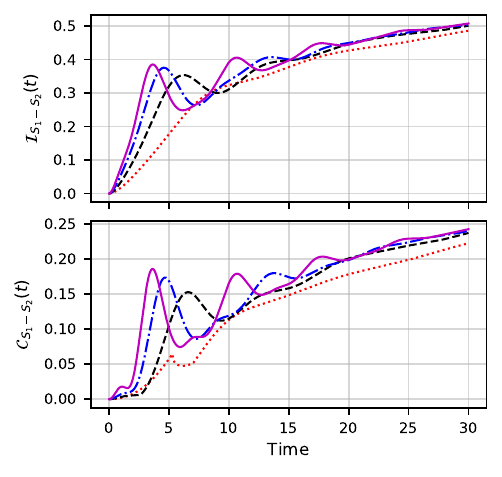}	
	}
	\hfill
	\subfloat[Cycle B ($T_{M_1}=0.8T_{M_2}$)\label{CORRELATIONCB}]{
		\includegraphics[width=1
		\columnwidth]{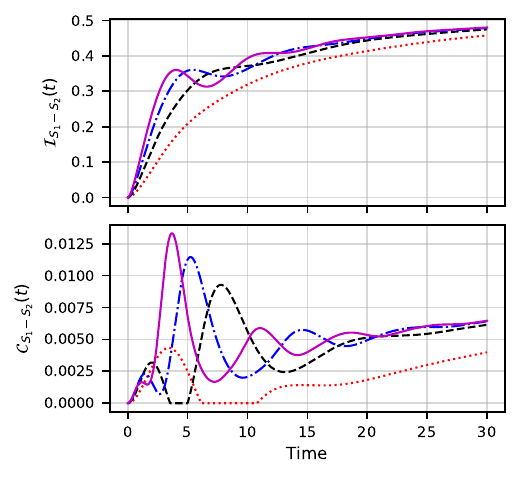}	
	}
\caption{Concurrence $\mathcal{C}_{S_1-S_2}(t)$ and mutual information $I_{S_1-S_2}(t)$ between the external system qubits. (a,b) Density plots as functions of $T_{M_1}$ and time; the white  dashed line separates cycles A and B and corresponds to the condition $T_{M_1}/T_{M_2}=0.5$. (c) Cycle A: significant entanglement generation for different coupling strengths. (d) Cycle B: weak or vanishing entanglement despite similar total correlations. The coupling strength is varied as $g=0.03,\,0.05,\,0.07,$ and $0.09$ (in units of $E_{M_2}$), shown by red (dotted), black (dashed), blue (dot--dashed), and magenta (solid) curves.}
	\label{ENTANGLMENT}	
\end{figure*}
%%%%%%%%%%%%%%%%%%%%%%%%%%%%%%%%%%%%%%%%%%%%%%%%%%%%%%%%%%%%%%%%%%%%%%%%%%%%%%%%%%%%%%%%%%%

In Fig.~\ref{ENTANGLMENT}, we analyze the entanglement and correlations in the external system $S$ over time. We set the coupling $g = 0.9$, and we plot $\mathcal{C}_{S_1 - S_2}(t)$ and $\mathcal{I}_{S_1 - S_2}(t)$ as functions of time and temperature $T_{M_1} \in [0.1 T_{M_2}, T_{M_2}]$, to illustrate the effect of the QATM $M$ cycles A and B on entanglement and coherence generation in Fig.~\ref{CONCURANCETEMPERATURES} and Fig.~\ref{CORRELATION}, respectively. 

We observe that when the temperature is below the white dashed line ($T_{M_1} < 0.5 T_{M_2}$, cycle A), the concurrence increases as the temperature of qubit $M_1$ approaches $T_{M_1} = 0.1 T_{M_2}$. Conversely, when the temperature is above the white dashed line ($T_{M_1} \geq 0.5 T_{M_2}$, cycle B), the concurrence decreases and approaches zero, remaining much smaller than in cycle A. In contrast, the mutual information is symmetric with respect to the white dashed line ($T_{M_1} = 0.5 T_{M_2}$, the transition point between cycles A and B). Physically, this indicates that when the QATM $M$ operates in cycle A, entanglement is generated, whereas the total correlation remains similar in both cycles A and B.

In Fig.~\ref{CORRELATIONCA} and Fig.~\ref{CORRELATIONCB}, we examine the effect of the coupling $g$ on the dynamics of entanglement and correlation in the external system $S$ for cycles A and B, respectively. As mentioned earlier, the total correlation is similar for both cycles A and B across all values of $g$. Regarding entanglement, we observe that in cycle A, entanglement is generated over time, while in cycle B, the concurrence remains small compared to cycle A. For any coupling value $g$, the concurrence dynamics are similar in magnitude but differ in their monotonic evolution, as $g$ governs the non-Markovian dynamics in our theoretical model.
\subsection{Role of coherence correlation on entanglement generation}
\label{sec:Role of coherence correlation on entanglement generation}
In this part, we show the role of coherence correlation noted $\Delta C_{S}(t)$, and the local coherences of the external system qubit on the entanglement generation, for the two cycles A and B of the QATM $M$. We use the relative entropy of coherence, denoted $C_{S_i}(t)$ for $i = \{1, 2\}$, for the qubits $S_1$ and $S_2$, respectively, and $C_{S}(t)$ as the global relative entropy of coherence of $S$, which are mathematically defined as:

\begin{align}
	C_{S}(t)=&S(\bar{\rho}_{S}(t))-S(\hat{\rho}_{S}(t)),\\
	C_{S_i}(t)=&S(\bar{\rho}_{S_i}(t))-S(\hat{\rho}_{S_i}(t)),\quad i=\{1,2\},\\
	\Delta C_{S}(t)=& C_{S}(t)- \sum_{i=1}^{2}C_{S_i}(t),
\end{align}

where $S(\bar{\rho}_{S}(t))$ and $S(\bar{\rho}_{S_i}(t))$ are the von Neumann entropies of the fully dephased states of the global system $S$ and the local systems $S_i$ for $i = \{1, 2\}$.

%%%%%%%%%%%%%%%%%%%%%%%%%%%%Fig_7%%%%%%%%%%%%%%%%%%%%%%%%%%%%%%%%%%%%%%%%%%%%%
\begin{figure*}[ht!]
	\subfloat[Cycle A\label{COHERENCECYCLEA}]{
		\includegraphics[width=1
		\columnwidth]{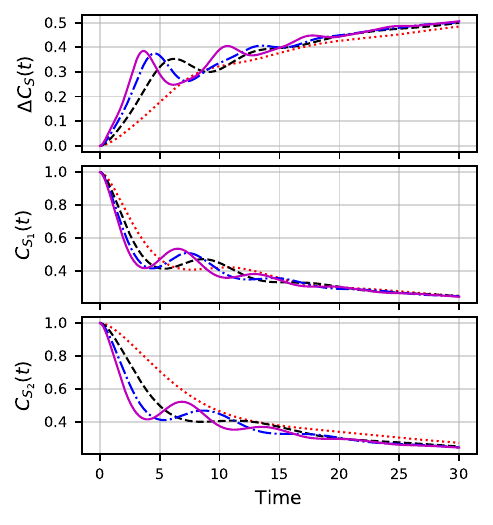}
	} \hfill
	\subfloat[Cycle B\label{COHERENCECYCLEB}]{
		\includegraphics[width=1
		\columnwidth]{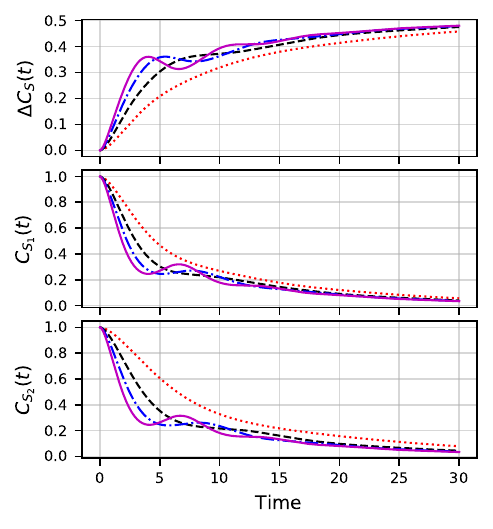}	
	}
	\caption{Local coherences $C_{S_1}(t)$, $C_{S_2}(t)$ and coherence correlation $\Delta C_S(t)$ in the external system.  
		(a) Cycle A: non-monotonic decay of local coherences with persistent and high coherence correlation.  
		(b) Cycle B: faster decay of local coherences and lower coherence correlation, consistent with weaker non-Markovianity. The coupling strength is varied as $g=0.03,\,0.05,\,0.07,$ and $0.09$ (in units of $E_{M_2}$), represented by red (dotted), black (dashed), blue (dot--dashed), and magenta (solid) curves.}
	\label{COHERENCE}	
\end{figure*}
%%%%%%%%%%%%%%%%%%%%%%%%%%%%%%%%%%%%%%%%%%%%%%%%%%%%%%%%%%%%%%%%%%%%%%%%%%%%%%%%%%%%%%%%%%%
In Fig.~\ref{COHERENCE}, we analyze the role of coherence correlations $\Delta C_S(t)$ and the local coherences $C_{S_i}(t)$ for $i \in \{1, 2\}$ as functions of time and the coupling between the QATM and the external system, for cycles A and B, shown in Fig.~\ref{COHERENCECYCLEA} and Fig.~\ref{COHERENCECYCLEB}, respectively. We observe that for both cycles, the local coherences of the qubits $S_1$ and $S_2$ decrease non-monotonically, indicating the presence of memory effects during the coherence evolution. Moreover, the local coherence of the external subsystems in cycle B decays to zero more quickly than in cycle A, due to the stronger non-Markovian effects present in cycle A.

For the coherence correlation $\Delta C_S(t)$, we observe a non-monotonic increase over time for both cycles, implying that the global coherence of the total external system $S$ is larger than the sum of the local coherences of $S_1$ and $S_2$, i.e.,
\begin{equation}
	C_{S}(t) > \sum_{i=1}^{2} C_{S_i}(t),
\end{equation}
which arises from the exchange of coherence correlations between the qubits $S_1$ and $S_2$ over time. Physically, there is a relationship between the total correlations (classical and quantum), quantified by the mutual information $\mathcal{I}_{S_1 - S_2}(t)$, and the coherence correlation $\Delta C_{S}(t)$, given mathematically by
\begin{align}
	\Delta C_{S}(t) = \mathcal{I}_{S_1 - S_2}(t) - \bar{\mathcal{I}}_{S_1 - S_2}(t),
\end{align}
where $\bar{\mathcal{I}}_{S_1 - S_2}(t) = \sum_{i=1}^{2} S(\bar{\rho}_{S_i}(t)) - S(\bar{\rho}_{S}(t))$ denotes the classical mutual information between $S_1$ and $S_2$, computed from the fully dephased states. 

This clearly illustrates the role of coherence correlations in the creation of quantum correlations between $S_1$ and $S_2$ over time. Moreover, if the external subsystems $S_1$ and $S_2$ are incoherent, we have $\mathcal{I}_{S_1 - S_2}(t) = \bar{\mathcal{I}}_{S_1 - S_2}(t)$, implying that the coherence correlation $\Delta C_S(t) = 0$, and consequently, the concurrence is also zero. This highlights that entanglement is directly linked to coherence correlations and confirms that the external qubits were initially prepared in superposition states.

In the context of open quantum systems, the local coherence of the external system qubits $S_1$ and $S_2$ decreases and eventually vanishes due to the decoherence effects created from the contact with reservoirs $R_1$ and $R_2$. However, correlations induced by the interaction between QATM and the external system allow $S_1$ and $S_2$ to become indirectly coupled through the thermal machine qubits $M_1$ and $M_2$. This interaction generates correlations over time between $S_1$ and $S_2$, which can be quantified using the mutual information. Physically, this means that the local coherences of the individual qubits are converted into coherence correlations, creating a correlated state for the external system and explaining the generation of entanglement. Hence, by comparing cycles $A$ and $B$, the loss of local coherence in cycle $B$ occurs more rapidly than in cycle $A$. This is because the dynamics in cycle $B$ are more irreversible, as reflected in the entropy production. Consequently, the correlation exchange between the external system and the QATM shows lower efficiency in cycle $B$. In contrast, memory effects arising from correlation exchange in cycle $A$ allow for mitigating the decoherence effects of the reservoirs more effectively.

\section{ Discussion and Conclusion}
This work investigates the relationship between quantum thermodynamics and quantum information theory in the framework of autonomous quantum thermal machines, with particular emphasis on entanglement generation and non-Markovian dynamics. We analyzed the influence of a quantum autonomous thermal machine on entanglement generation by considering the Hilbert space structure, temperature constraints, correlation exchange, memory effects, and correlation coherence as a quantum resource. In particular, we examined how the virtual temperature and the energy spacing of the QATM and external system qubits determine the realization of two thermodynamic cycles, denoted $A$ and $B$, and how these cycles affect non-Markovianity and entanglement generation.

The considered quantum autonomous thermal machine (QATM) consists of two qubits, $M_1$ and $M_2$, each coupled to its own bosonic Markovian reservoir, $R_1$ and $R_2$, at different temperatures. The external system $S$ is composed of two uncoupled qubits, $S_1$ and $S_2$. By analyzing the Hilbert space structure and employing the concept of virtual temperature, we established a common interaction between the QATM and the external system. Temperature and energy constraints were derived in order to realize two thermodynamic cycles of the QATM. The transition between cycles $A$ and $B$ is determined by the relation $T_{M_1} = \frac{E_{M_2}}{E_{M_1}} T_{M_2}$.

The robustness of the theoretical model was examined numerically through the analysis of heat exchange between the QATM qubits ($M_1$, $M_2$) and the external system qubits ($S_1$, $S_2$), as well as through the evolution of the reservoir temperatures $T_{M_1}$ and $T_{M_2}$. The transition from cycle $A$ to cycle $B$ is characterized by the transition temperature $T_{M_1} = \frac{E_{M_2}}{E_{M_1}} T_{M_2}$. The second law of quantum thermodynamics was analyzed via entropy production, showing that the irreversibility associated with cycle $A$ is smaller than that of cycle $B$. This result indicates that cycle $A$ suppresses reservoir-induced decoherence more effectively.

The entropy production rate can take negative values during the evolution, which indicates the presence of memory effects. We quantified non-Markovianity and found that it is more pronounced in cycle $A$ than in cycle $B$, particularly for the external system qubits $S_1$ and $S_2$. This behavior arises from the exchange of correlations, where the QATM effectively acts as a structured non-Markovian reservoir for the external system.

Entanglement is generated in the external system only when the QATM operates in cycle $A$, which is characterized by lower irreversibility and stronger non-Markovianity compared with cycle $B$. In contrast, the total correlations remain similar in both cycles since they are associated with the global properties of the external system. Cycle $A$ therefore provides a more effective mechanism for controlling reservoir-induced decoherence in the external system. Furthermore, correlation coherence plays a crucial role as a resource for entanglement generation in $S$, since incoherent states of the external system correspond to vanishing correlation coherence.

In relation to previous works, Khandelwal \textit{et al.}~\cite{INTRO7} and Brask \textit{et al.}~\cite{INTRO8} investigated entanglement generation using QATMs mainly in the steady-state regime, without considering external systems or memory effects. Aguilar \textit{et al.}~\cite{INTRO9} analyzed entanglement generation in thermal machines operating in different environments, but without addressing memory effects in autonomous thermal machines. In contrast, our model considers a QATM operating out of equilibrium and its influence on an external system through memory effects and Hilbert space structure, allowing the realization of thermodynamic cycles and extending the framework introduced in Ref.~\cite{INTRO10} beyond the steady-state regime.

Finally, the proposed theoretical framework is compatible with current experimental platforms based on superconducting qubits, since the considered parameters lie within the regimes explored in existing experimental implementations.

\appendix
\section{Autonomous Operation}
\label{Autonomous Operation}
In this appendix, we shall give the main details about the autonomous operation used in our work. Indeed, assume that the total system evolves according to  the following density matrix: 
\begin{equation}
	\hat{\rho}(t) = \hat{U}(t)\hat{\rho}(0)\hat{U}^{\dagger}(t).
\end{equation}
Then, the total energy, namely $E(t) = \mathrm{Tr}\{\hat{\rho}(t)\hat{H}\}$ is conserved in time, that is, $\Delta E(t)=0$. This is consistent with the first law of quantum thermodynamics~\cite{INTRO1,INTRO2,INTRO3}.  However, for the subsystems $M$ and $S$, the internal energy is given as: 
\begin{equation}
	E_{MS}(t) = \mathrm{Tr}\{(\hat{H}_{S} + \hat{H}_{M})\hat{\rho}_{MS}(t)\}.
\end{equation}
According to the first law of quantum thermodynamics, the variation of internal energy can be decomposed into heat and work contributions as follows: 
\begin{eqnarray}
	\Delta E_{MS}(t) &=& Q_{MS}(t) + W_{MS}(t), \nonumber \\
Q_{MS}(t) &=& \mathrm{Tr}_{M,S}\{(\hat{H}_{S} + \hat{H}_{M})\,\Delta\hat{\rho}_{MS}(t)\}, \nonumber \\
W_{MS}(t) &=& \mathrm{Tr}_{M,S}\{\hat{\rho}_{MS}(t)\,(\Delta\hat{H}_{S} + \Delta\hat{H}_{M})\},
\end{eqnarray}
where $Q_{MS}(t)$ and $W_{MS}(t)$ denote heat and work exchanged by the subsystems $M$ and $S$, respectively~\cite{INTRO1,INTRO2,INTRO3}.  
In our scenario, the Hamiltonians $\hat{H}_S$ and $\hat{H}_M$ are time-independent, meaning that: 
\begin{equation}
	\Delta \hat{H}_S = \Delta \hat{H}_M = \hat{0}.
\end{equation}
Physically, this means that no external time-dependent driving is applied to the system. Consequently, no work is exchanged during the evolution, $W_{MS}(t)=0$, and the subsystems evolve autonomously.  
\section{Virtual temperature}
\label{Virtual temperature}
The QATM $M$ and external system $S$ can be treated as a single virtual qubit with energy spacing $E_M = E_S = E_{M_2} - E_{M_1}$, where the ground and excited states are
\[
\ket{0}_{M(S)} = \ket{1_{M_1(S_1)}0_{M_2(S_2)}}, \quad
\ket{1}_{M(S)} = \ket{0_{M_1(S_1)}1_{M_2(S_2)}}.
\]
The populations of these states, $P_{M(S)}^{G}(t)$ and $P_{M(S)}^{E}(t)$, are defined as bellow: 
\begin{eqnarray}
	P_{M(S)}^{G}(t) &=& P_{M_1(S_1)}^{E}(t) P_{M_2(S_2)}^{G}(t), \nonumber\\
	P_{M(S)}^{E}(t) &=& P_{M_1(S_1)}^{G}(t) P_{M_2(S_2)}^{E}(t),
\end{eqnarray}
where $P_{M_i(S_i)}^{G}(t) = \braket{0_{M_i(S_i)}|\hat{\rho}_{M_i(S_i)}(t)|0_{M_i(S_i)}}$ and $P_{M_i(S_i)}^{E}(t) = \braket{1_{M_i(S_i)}|\hat{\rho}_{M_i(S_i)}(t)|1_{M_i(S_i)}}$ are the ground and excited state populations of qubit $M_i(S_i)$ for $i = \{1,2\}$.  
However, since the QATM qubits follow Boltzmann-like distributions characterized by effective instantaneous temperatures \cite{R8}, one can write
	\begin{align}
		P_{M_i}^{E}(t) &= P_{M_i}^{G}(t)\, e^{-E_{M_i}/T_{M_i}}, \quad i=\{1,2\}, \nonumber\\
		P_{M}^{E}(t) &= P_{M}^{G}(t)\, e^{-E_{M}/T_{M}},
		\label{Popm}
\end{align}
where $E_M$ is the energy spacing of the virtual qubit and $T_M$ is the corresponding virtual temperature defined using the following form: 
\begin{align}
	T_M = \frac{E_M}{\frac{E_{M_2}}{T_{M_2}} - \frac{E_{M_1}}{T_{M_1}}}. \label{Virt}
\end{align}
Note that at equilibrium, $T_M$ satisfies $T_M = T_{M_1} = T_{M_2}$. But in general $T_M$ determines the temperature constraints distinguishing Cycles $A$ and $B$.
%----------------------------------------------------------------------------------------------------------

\section*{Acknowledgment}

A.~K acknowledges CNRST-Morocco for financial support through the program ``PhD-ASsociate Scholarship – PASS''. 
A.~K also acknowledges the hospitality of Prof.~Özgür E. Müstecaplıoğlu and his research group at the Department of Physics, Koç University, Istanbul, Sarıyer 34450, Türkiye, during a scientific visit where part of this work was carried out. A.E.A. completed part of this work during a research visit to the Laboratoire de Physique et Modélisation des Milieux Condensés (LPMMC), CNRS, in Grenoble, France. He extends his sincere gratitude to the CNRS - Fédération de Recherche QuantAlps, Comité de Direction QuantAlps, for their financial support and for fostering a stimulating and friendly research environment. 

\section*{Declaration of Interest}
The authors declare that they have no conflict of interest.
\section*{Data availability statement}
No data statement is available.

\end{document}